\documentclass[prd,preprintnumbers,superscriptaddress,amsmath,amssymb,nofootinbib,twocolumn,floatfix]{revtex4-2}

\usepackage{setspace}
\usepackage{graphicx}
\usepackage{hyperref}
\usepackage{booktabs}    
\usepackage{units}
\usepackage{nameref}

\usepackage{xcolor}

\newcommand{\sectionname}[1]{ \noindent{\bfseries #1}.---}

\newcommand{\be}{\begin{equation}}
\newcommand{\ee}{\end{equation}}

\usepackage{soul}

\begin{document}

\preprint{KCL-PH-TH-2025-16}
\title{Gravitational-wave background detection using machine learning}

\author{Hugo Einsle}
\email{hugo.einsle@oca.eu}
\affiliation{Universit\'e C\^ote d'Azur, Observatoire de la C\^ote d'Azur, CNRS, Artemis, Nice 06300, France}
\author{Marie Anne Bizouard}
\email{marieanne.bizouard@oca.eu}
\affiliation{Universit\'e C\^ote d'Azur, Observatoire de la C\^ote d'Azur, CNRS, Artemis, Nice 06300, France}
\author{Tania Regimbau}
\email{regimbau@lapp.in2p3.fr}
\affiliation{LAPP, CNRS, 9 Chemin de Bellevue, 74941 Annecy-le-Vieux, France}
\author{Mairi Sakellariadou}
\email{mairi.sakellariadou@kcl.ac.uk}
\affiliation{Theoretical Particle Physics and Cosmology Group, \, Physics \, Department, \\ King's College London,  University of London,  Strand,  London, WC2R  2LS, UK}
\affiliation{Universit\'e C\^ote d'Azur, Observatoire de la C\^ote d'Azur, CNRS, Artemis, Nice 06300, France}

\date{\today}
\begin{abstract}
Extracting the faint gravitational-wave background (GWB) signal from dominant detector noise and disentangling its 
astrophysical and cosmological components remain significant challenges for traditional methods like cross-correlation analysis. 
We propose a novel hybrid approach that combines deep learning with Bayesian inference to identify and characterize the GWB more rapidly than current techniques. Our method utilizes a custom-designed multi-scale multi-headed autoencoder (MSMHAutoencoder) architecture 
to separate GWB signals from detector noise, and subsequently Marcov Chain Monte Carlo parameter estimation 
to disentangle the GWB components.
Using simulated data representative of the LIGO-Virgo-KAGRA network at design sensitivity, we show that our MSMHAutoencoder can detect with high confidence (log noise Bayes factor of 3) a GWB from binary black hole mergers with fractional energy density $\Omega_{\text{BBH}} \approx 10^{-9}$ at 25 Hz. 
In the presence of such an astrophysical GWB,
we can simultaneously measure a cosmological 
component as faint as $\Omega_{\text{Cosmo}} \approx 1.3 \times 10^{-10}$ 
using 47.4 days of training data. 
\end{abstract}
\maketitle

\section{Introduction}

The LIGO-Virgo-KAGRA (LVK) Collaboration has announced 90 transient gravitational-wave signals, resulting from compact
binary coalescences, detected during the first three observing runs (O1-O3)~\cite{LIGOScientific:2018mvr,LIGOScientific:2020ibl,GWTC-3-LIGOScientific:2021djp}. Currently, the LVK Collaboration has released about 200 public alerts from the fourth observing run (O4) ~\cite{Capote_2025}.
Apart from the detected transient gravitational wave signals, the Universe is expected to be permeated by a gravitational-wave background (GWB) ~\cite{Christensen:2018iqi}  from cosmological~\cite{Caprini:2018mtu,Kalogera:2021bya,LISACosmologyWorkingGroup:2022jok} and astrophysical~\cite{Regimbau:2011rp} origin. The former results from violent early Universe processes, such as first-order phase transitions, topological defects (domain walls, cosmic strings), or a period of inflation. 
The latter is due to the superposition of gravitational waves from weak or distant astrophysical sources, such as stellar black hole and neutron star mergers, rotating or oscillating neutron stars, core collapses, among others.

The LVK Collaboration is actively searching for a GWB, as it presents a powerful and unique tool to constrain high-energy physics models beyond the Standard Model at energy scales unreachable by accelerators~\cite{LIGOScientific:2021nrg,Caldwell_2022,Blasi:2022ayo}, and early Universe scenario~\cite{Martinovic:2021hzy,Badger:2022nwo,Duval:2024jsg,Badger:2024ekb}. The detection of a GWB of astrophysical origin offers a unique window into the average properties of compact binary populations, particularly at high redshifts that are inaccessible to individual source detections. Such a background can reveal valuable astrophysical information, including the mass distribution of neutron star and black hole progenitors, as well as the merger rates of compact binaries across cosmic time~\cite{LIGOScientific:2017zlf}.

Detecting a GWB is a challenging undertaking as the noise (detected source foreground, environmental, or instrumental) dominates the signal~\cite{Christensen:1996da}.
The method followed by the LVK Collaboration is based on cross-correlating
the strain data between pairs of gravitational-wave detectors~\cite{LIGOScientific:2016jlg, LIGOScientific:2019vic, KAGRA:2021kbb}. Cross-correlation is preferred to auto-correlation methods as the noise variances in each detector are not known
sufficiently well to allow subtraction of the noise auto-power. The presence of correlated noise between the detectors can be subtracted following methods as such discussed in~\cite{Meyers:2020qrb}.

Once a successful GWB detection is made, the challenge of untangling the signals to identify the contributing sources will be even greater.
Traditional statistical methods, such as parameter estimation via Markov Chain Monte Carlo (MCMC), are often hindered by the high dimensionality and multimodal nature of the posterior distributions ~\cite{Parida_2016, Suresh_2021, Boileau_2021}. 
Hence, disentangling the GWB into its constituent requires innovative approaches beyond conventional analysis.

Deep learning has recently emerged as a promising alternative for gravitational-wave data analysis. Architectures such as convolutional neural networks and autoencoders have demonstrated exceptional capabilities in extracting complex signals from high-dimensional data across various fields~\cite{Carleo:2019ptp,ntampaka2019deep}. In gravitational-wave astronomy, these techniques have enabled advances in signal detection~\cite{George:2017pmj,Schafer:2022dxv}, waveform modeling~\cite{Chua:2019wwt}, and parameter estimation~\cite{Gabbard:2019rde} (see references in a recent review~\cite{Cuoco:2024cdk}). Leveraging these successes, the application of deep learning to GWB analysis offers new possibilities to overcome the limitations of traditional methods. 

In what follows, we present a novel mixed deep-learning Bayesian inference approach aiming firstly at identifying GWB in the presence of detector noise and secondly disentangling its components (astrophysical and cosmological), quicker than what it is expected with the cross-correlation method, currently used by the LVK Collaboration. Our approach utilizes a custom-designed multi-scale multi-headed autoencoder architecture specifically tailored to separate the overlapping contributions of astrophysical and cosmological sources, while mitigating the effects of the significantly louder intrinsic detector noise. To create simulated data, we model detector noise as Gaussian and uncorrelated between the network detectors. The astrophysical GWB is generated from populations of compact binary coalescences, while the cosmological one is simulated using power spectral density modeled as a power law. By training the network on realistic noise conditions and through targeted injections, our autoencoder architecture learns to separate the GWB signal from detector noise.
The separated output then serves as the basis for subsequent MCMC-based joint parameter estimation, allowing to assess the individual contributions of the different GWB components. To validate our hybrid approach -- machine learning combined with parameter estimation techniques -- we investigate its performance, considering first a single astrophysical GWB, followed by the case where a cosmological GWB is also present.

The paper is organized as follows:
In Section~\ref{sec: state of the art}, we briefly present the current LVK Collaboration method to detect a GWB based on cross-correlations, and highlight approaches discussed in the literature for disentangling its components.
In Section~\ref{sec:nn_arch_training}, we present the neural network architecture and the training methodology; details are given in the appendices.
In Section~\ref{sec: Data Model}, we discuss the simulated astrophysical and cosmological GWB data, and the detector noise.
In Section~\ref{sec:validation}, we demonstrate the validation of our deep learning approach. We wrap up our conclusions in Section~\ref{sec: Conclusions}. 

\section{State of the art of GWB measurement}
\label{sec: state of the art}

The strength of the isotropic GWB is usually described in terms of a dimensionless energy-density spectrum
\begin{equation}
\Omega_{\rm GW}(f) = \frac{f}{\rho_{\rm c}} \frac{{\rm d}\rho_{\rm GW}}{{\rm d} f}~,
\end{equation}
which quantifies the gravitational-wave energy density ${\rm d}\rho_{\rm GW}$ per logarithmic frequency interval, normalized by the critical energy density $\rho_{\rm c} = 3H_0^2 c^2 / (8\pi G)$. It is related to the gravitational-wave strain power spectral density $S_{\rm h}(f)$ by
\begin{equation}
S_{\rm h}(f) = \frac{3H_0^2}{2\pi^2} \frac{\Omega_{\rm GW}(f)}{f^3}~.
\end{equation}
A common phenomenological model assumes a power-law form
\begin{equation}
\Omega_{\rm GW}(f) = \Omega_\alpha \left(\frac{f}{f_{\rm ref}}\right)^\alpha~,
\end{equation}
where $\Omega_\alpha$ is the amplitude at a reference frequency $f_{\rm ref}$ and $\alpha$ is the spectral index. For unresolved compact binary coalescences (CBCs) from Population I/II stars, the inspiral-dominated portion of the background leads to an approximate power-law with $\alpha = 2/3$, while rotating neutron stars with ellipticity, including magnetars, emit a background with a steep spectrum scaling as $\alpha = 4$ (see~\cite{Regimbau:2011rp} for a review and references therein). The spectrum from core-collapse supernovae is highly uncertain and generally not well approximated by a simple power-law. In some population-based models, a steep slope (e.g., $\alpha \sim 3$) has been proposed, but this depends strongly on assumptions and is more relevant to the sub-hertz band targeted by space-based detectors~\cite{2005PhRvD..72h4001B}. Cosmological sources such as slow-roll inflation and cosmic strings (over the frequency range relevant for LVK~\cite{Auclair:2019wcv}) typically produce a nearly scale-invariant spectrum with $\alpha \simeq 0$ . First-order phase transitions generate a broken power-law spectrum with a peak frequency determined by the energy scale and duration of the transition~\cite{Caprini:2018mtu}. 

The detection of a 
gravitational-wave background relies on statistical analysis of the data collected by multiple interferometers. The LVK Collaboration, for example, employs a cross-correlation technique under the assumption of a Gaussian, stationary, unpolarized, and isotropic background.
To cross-correlate data of interferometers $i, j$ we built the estimator
\be
\hat C_{i,j}(f;t)=\frac{2}{T}\frac{{\rm Re}[\tilde s^\star_i(f;t)\tilde s_j(f;t)]}{\gamma_{ij}(f)S_0(f)}~,
\ee
where $\tilde s^\star_i(f;t)$ is the Fourier transform of the strain time
series in interferometer $i$ starting at time $t$, $T$ is the segment duration to compute the Fourier transform, $S_0(f)$ $S_0(f)=3H_0^2/(10\pi^2 f^3)$ and 
$\gamma_{ij}(f)$ is the normalized overlap reduction function between interferometers $i, j$, which quantifies how the detectors’ relative positions and orientations reduce the correlated response to a common GWB signal. For tensor polarizations it is given
by~\cite{Christensen:1996da} 
\begin{equation}
\gamma_{ij}(f) = \frac{5}{8\pi} \int_{S^2} d\hat{\Omega} \, e^{2\pi i f \hat{\Omega} \cdot \Delta \mathbf{x} / c} \sum_{A=+,\times} F_1^A(\hat{\Omega}) F_2^A(\hat{\Omega})~,
\end{equation}
where \( \hat{\Omega} \) is the direction of gravitational wave propagation, \( \Delta \mathbf{x} \) is the separation vector between the detectors, \( c \) is the speed of light, and \( F_j^A(\hat{\Omega}) \) are the antenna pattern functions of detector \( j \) for the two polarizations \( A = +, \times \).

Assuming the gravitational-wave signal and the intrinsic noise are uncorrelated, and that the noise in each frequency bin is independent, one can show that
\be
\langle \hat C_{ij}(f;t)\rangle = \Omega_{\rm GW}(f)+
\frac{2}{T}
\frac{{\rm Re}\left[\langle \tilde n_i^\star(f;t)\tilde n_j(f;t) \rangle\right]}
{\gamma_{ij}(f)S_0(f)}~.
\ee
In the absence of correlated noise, $\langle \hat C_{ij}(f;t)\rangle =\Omega_{\rm GW}(f)$, we have built an estimator of the GW energy density. Otherwise, we have to estimate the contribution of
correlated noise to the GWB estimator~\cite{Martinovic:2021hzy} or perform Wiener filtering~\cite{Thrane:2014yza}.

The successful measurement of the GWB will lead to the next critical challenge: disentangling its various components, particularly distinguishing between astrophysical and cosmological contributions. In~\cite{Martinovic:2020hru}, the authors model the GWB as a combination of compact binary coalescences and cosmological sources (such as cosmic strings and first-order phase transitions). They apply Bayesian parameter estimation and model selection techniques to statistically separate these components based on their spectral properties. 
Their results 
show that the Advanced LIGO and Advanced Virgo network, operating at design sensitivity and including instrumental noise, does not have the capability to distinguish the cosmological signal from the dominant astrophysical background. 
For the latter they consider a uniformly distributed CBC background with $\Omega_{\alpha=2/3}\in (10^{-9.4}, 10^{-8.4})$.
However, they also demonstrate that a third-generation detector network could detect cosmological signals as weak as $\Omega_{\rm CS}(25~ {\rm Hz}) = 4.5\times 10^{-13}$ (from cosmic strings) and $\Omega_{\rm FOPT}(25 ~{\rm Hz}) = 2.2\times 10^{-13}$ (from first-order phase transitions), assuming both components remain unresolved and are modeled statistically.

A complementary approach involves the proposed subtraction of individually resolved astrophysical sources to suppress the foreground and potentially reveal a residual cosmological background. This method has been explored in the context of third-generation detector networks such as the Einstein Telescope and Cosmic Explorer~\cite{2017PhRvL.118o1105R,2020PhRvD.102b4051S,2020PhRvD.102f3009S,2023PhRvD.108f4040Z}.
A related mitigation technique is introduced in~\cite{Zhong:2024dss}, which applies a time-frequency domain notching procedure to suppress parts of the astrophysical background that contaminate the signal. Their study, conducted in the context of third-generation detectors, finds that even with such notching, the unresolvable astrophysical background from neutron star binaries remains the primary barrier to detecting a cosmological signal. This confirms earlier results reported in~\cite{2020PhRvD.102b4051S}.

An alternative strategy is proposed in~\cite{Biscoveanu:2020gds}, where a Bayesian inference framework is used to estimate the cosmological GWB (with spectral index $\alpha = 0$) in the presence of an unresolved astrophysical foreground and instrumental noise, without subtracting resolved sources.

\section{Neural network architecture and training}
\label{sec:nn_arch_training} 
 
Let us highlight the specifically designed neural network autoencoder architecture~\cite{Bourlard1988, Hinton2006, Goodfellow-et-al-2016} we develop for separating GWB from the dominant 
noise of the gravitational-wave detectors. The neural network processes information across multiple frequency scales to capture diverse spectral features, and exploit the temporal stationary nature of spectra to distinguish GWB from transient noise. It employs a sequential estimation pathway to characterize the dominant noise component, and then incorporates a learned correction mechanism to isolate weak signals without resorting to simple subtraction which could destroy faint signatures. 

In what follows, we first briefly describe the model architecture, then outline the physics-informed training methodology, and finally highlight the curriculum learning strategy adapted for the low signal-to-noise ratio regime of the GWB.

\subsection{Model Architecture}
\label{subsec:model_arch}

\begin{figure*}[t]
    \centering
    \includegraphics[width=0.8\textwidth]{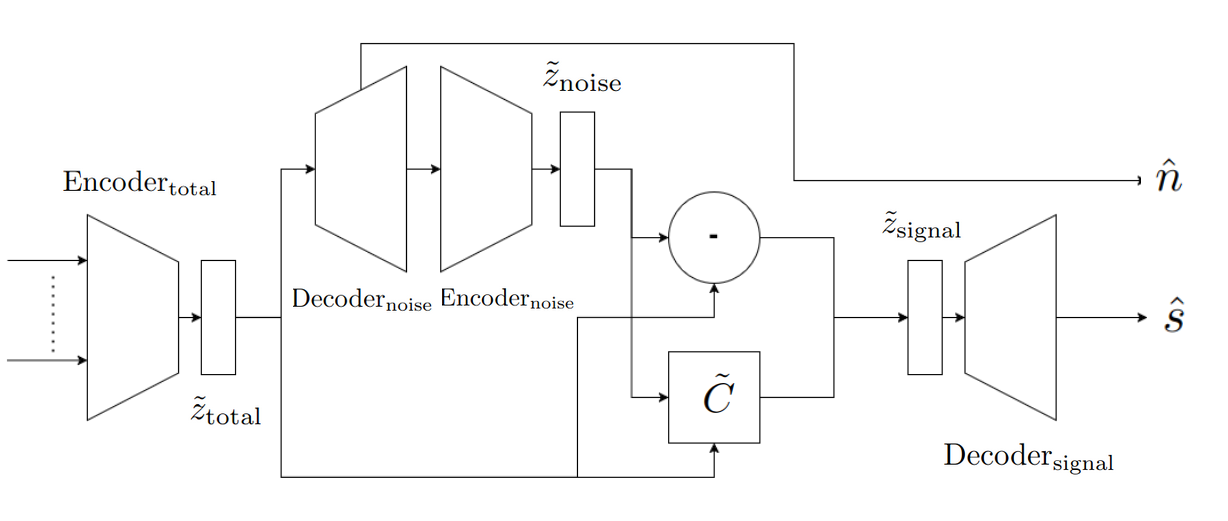} 
    \caption{
        Schematic representation of the MSMHAutoencoder architecture. The model processes frequency-domain inputs containing both noise and 
        signals. The encoder extracts hierarchical latent representations, which are used to estimate noise components via a dedicated decoder and subsequent re-encoding. Corrected latent spaces enable adaptive signal reconstruction, yielding separate outputs for noise and signal components. This design leverages multi-scale features and skip connections to ensure precise separation of the GWB from detector noise.
    }
    \label{fig:model_architecture}
\end{figure*}

Our deep learning algorithm, based on a custom-designed multi-scale multi-headed autoencoder architecture (hereafter called MSMHAutoencoder), is designed to separate GWB from 
detector noise. Structurally, an autoencoder features two neural network sections connected by a bottleneck layer. The first section, the encoder, reduces the dimensionality of the input data.
The data at the bottleneck 
contains the essential information needed for reconstruction. The second section, the decoder, 
decompresses the data in a sequence of layers aiming to reconstruct the output. We represent schematically the MSMHAutoencoder architecture in Figure~\ref{fig:model_architecture}. 
The input data $x \in \mathbb{R}^{M \times N}$ consist of $M$ consecutive spectra, each comprising $N$ frequency bins. While the instrumental noise in each of these $M$ spectra represents a different realization, 
the underlying GWB signal is assumed to be statistically stationary or to vary slowly over the corresponding timescale. 
Therefore, the signal component within each spectrum is treated as a distinct realization drawn from the same signal process. 
By jointly analyzing this sequence of $M$ spectra, the architecture can leverage the statistical consistency of the GWB signal 
to distinguish it from the noise components and transient artifacts.

Let us briefly summarize the various steps (and refer the reader to Appendix~\ref{app:architecture_details} for further details).
To process the input sequence $x$ and extract features to capture both the noise and the signal, the architecture uses Encoder$_{\text{total}}$. 
It processes the $M$ input spectra and constructs a latent representation, 
$\tilde{z}_{\text{total}}$ \footnote{The tilde notation (e.g., on $\tilde{z}$) signifies the complete set of multi-scale latent representations forming the effective bottleneck for an encoder or decoder pathway. This set is composed of latent tensors from $L$ different scales, formally defined as $\tilde{z} = \{z^\ell\}_{\ell=1}^L$, where each $z^\ell$ captures features at a specific scale $\ell$. Appendix~\ref{app:architecture_details} details this multi-scale structure and the properties of each $z^\ell$.}
\begin{equation}
{\rm Encoder}_{\text{total}}(x) \;\rightarrow\; \tilde{z}_{\text{total}}~,
\end{equation}
which constitutes the initial noise+signal latent space.
The latter is used by Decoder$_{\text{noise}}$ to reconstruct an estimate of the noise spectrum
\begin{equation}
{\rm Decoder}_{\text{noise}}( \tilde{z}_{\text{total}} ) \;\rightarrow\; \hat{n}~,
\end{equation}
where $\hat{n} \in \mathbb{R}^N$. To refine the separation between signal and noise,  
$\hat{n}$ is re-analyzed by a Encoder$_{\text{noise}}$ 
\begin{equation}
{\rm Encoder}_{\text{noise}}\bigl(\hat{n}\bigr) \;\rightarrow\; \tilde{z}_{\text{noise}}~.
\end{equation}
%
Separation of the signal occurs in the latent space. Based on empirical observations indicating that latent subtraction $\tilde{z}_{\text{total}} - \tilde{z}_{\text{noise}}$ alone does not reliably isolate the signal component, we introduce an additional learned correction mechanism. 
Using specialized corrector modules $\tilde{C}$, this mechanism processes features corresponding to the total input $\tilde{z}_{\text{total}}$ and the estimated noise $\tilde{z}_{\text{noise}}$, to produce adaptive refinements and generate signal-specific latent features
\begin{equation}
\tilde{z}_{\text{signal}} = (\tilde{z}_{\text{total}} - \tilde{z}_{\text{noise}}) + \tilde{C}(\tilde{z}_{\text{total}},\, \tilde{z}_{\text{noise}}),
\end{equation}
This adaptive correction allows the network to learn how to best isolate signal features present in the latent space.
Finally, 
Decoder$_{\text{signal}}$, processes the corrected signal latent features $\tilde{z}_{\text{signal}}$ to reconstruct the estimated GWB component
\begin{equation}
{\rm Decoder}_{\text{signal}} \;(\tilde{z}_{\text{signal}}) \;\rightarrow\; \hat{s}~,
\end{equation}
where $\hat{s} \in \mathbb{R}^{N}$ is the isolated GWB spectrum.

\subsection{Training Methodology}
\label{subsec:training_methodology}

We train the MSMHAutoencoder using a supervised methodology on simulated noisy GWB spectra for which the signal and noise components are known. The primary goal is to iteratively adjust the MSMHAutoencoder's neuron weights to accurately separate signals from detector noise. 
Since signals may span several orders of magnitude, we apply our model primarily on spectra in the logarithmic domain to effectively handle this wide dynamic range.

The training process minimizes a physics-informed loss function via {\sl backpropagation}~\cite{Rumelhart1986}. 
This procedure transmits the error from the output layer back through the network, layer by layer, allowing the model to adjust its internal weights, reducing the discrepancy between predictions and targets. We employ separate training and validation datasets to control the training against overfitting. The former inform neuron weights, while the latter which are not used during direct neuron weights optimization, provide an unbiased estimate of the model's performance. 
This implementation guides the dynamic adjustment of key hyperparameters, such as the learning rate $\eta$, and triggers an early stopping mechanism to halt training and prevent overfitting. 

The total loss function $\mathcal{L}_{\text{total}}$ is tailored 
to include accurately several components
\begin{equation}
\mathcal{L}_{\text{total}} = \lambda_s \mathcal{L}_{\text{spectral}} + \lambda_l \,\mathcal{L}_{\text{latent}} + \lambda_c \,\mathcal{L}_{\text{consist}}
\label{eq:total_loss}~,
\end{equation}
where $\mathcal{L}_{\text{spectral}}$ enforces reconstruction accuracy and spectral smoothness, $\mathcal{L}_{\text{latent}}$ promotes consistent internal representation of noise, and $\mathcal{L}_{\text{consist}}$ ensures self-consistency between input and 
reconstructed components. The positive or null weights $\lambda_s, \lambda_l, \lambda_c$ are hyperparameters that balance the contribution of these different components.

The core spectral loss, \( \mathcal{L}_{\text{spectral}} \), is computed on batches of training data. Each batch contains \( B \) sets of \( M \) 
pairs of input-output log-spectra, for 
signal and noise components, drawn from the training dataset. 

The function \( \mathcal{L}_{\text{spectral}} \) combines the reconstruction terms for signal and noise with a spectral {\sl smoothness penalty}. This penalty is designed to encourage smoother and more physically plausible predicted signal spectra by penalizing rapid variations or high-frequency oscillations. This is achieved by incorporating a second-order derivative penalty term, $\|\nabla^2 \hat{s}_i\|_2^2 $ into the loss function
\begin{equation}
\mathcal{L}_{\text{spectral}} = \frac{1}{B} \sum_{i=1}^{B} \Bigl[\|\hat{n}_i - n_i\|_2^2 + w(k_i) \,\|\hat{s}_i - s_i\|_2^2 \Bigr] + \|\nabla^2 \hat{s}_i\|_2^2~.
\end{equation}
with
\begin{equation}
\|\nabla^2 \hat{s}_i\|_2^2 = \sum_{j=1}^{N-2} \left[\hat{s}_{i,j+2} - 2\hat{s}_{i,j+1} + \hat{s}_{i,j}\right]^2~,
\end{equation}
where, as stated earlier, the index $j$ refers to the frequency.
For a given set \( i \), the vectors \( \hat{n}_i, n_i, \hat{s}_i, s_i \in \mathbb{R}^{N} \) represent the predicted and true log-spectra of the noise and signal components, and the notation \( \|~\|_2^2 \) refers to the squared Euclidean norm. 
During training, different combinations of  
signals are used 
and the reconstruction weight $w(k_i)$ depends on the 
category $k_i$ of the $i$-th sample,  
increasing the penalty for inaccurate signal predictions when 
noise is only present ($k_i = k_{\text{noise-only}}$)
\begin{equation}
w(k_i) = \begin{cases}
w_{\text{high}} & \text{if } k_i = k_{\text{noise-only}} \\
w_{\text{base}} & \text{otherwise}~.
\end{cases}
\end{equation}
Note that $w_{\text{high}} > w_{\text{base}} \ge 0$ are predefined weights
for the categories $k=0$ (Noise only), $k=1$ (Noise+Cosmo), $k=2$ (Noise+Astro), and $k=3$ (Noise+Cosmo+Astro). 
This adaptive weighting strategy is crucial for preventing the model from erroneously generating or identifying spurious signal-like features based on noise particularly in noise-only data segments.

The latent loss term $\mathcal{L}_{\text{latent}}$ compares the 
features produced by the noise encoder network Encoder$_{\text{noise}}$ when processing the predicted noise $\hat{n}$ versus the true one $n$. 
It reads
\begin{equation}
\mathcal{L}_{\text{latent}} = \| \tilde{z}_{\text{noise}}(\hat{n}) - \tilde{z}_{\text{noise}}(n) \|_F^2~,
\end{equation}
where 
$\tilde{z}_{\text{noise}}$ denotes the output from $\rm Encoder_{\text{noise}}$, and 
$\| ~ \|_F^2$ is the squared Frobenius norm, summing squared differences across all elements in the batch's latent representations.

Finally, $\mathcal{L}_{\text{consist}}$ 
ensures that the prediction matches the original data in the latent space.
To do so, the components are first recombined in the linear domain, 
converted back to the log domain\footnote{
$\hat{x} = \log_{10}(10^{\hat{n}} + 10^{\hat{s}} + \epsilon)$, with $\epsilon$ is a small constant used for numerical stability.},  and then re-encoded by $\rm Encoder_{\text{total}}$
\begin{equation}
\mathcal{L}_{\text{consist}} = \| \tilde{z}_{\text{total}}(\hat{x}) - \tilde{z}_{\text{total}}(x) \|_F^2~.
\end{equation}

The weights of the MSMHAutoencoder's neurons optimization is performed using the Adam algorithm ~\cite{Adam}. To prevent numerical instability, and in particular gradient explosion, the Euclidean norm of the gradients is restricted to a maximum value before neuron weight updates. The learning rate, starting at $\eta_0$, is adaptively managed using a scheduling strategy that reduces the rate based on validation loss stagnation over a defined {\sl patience period}. Training terminates via the early stopping mechanism~\cite{Prechelt_1998_EarlyStopping, Goodfellow-et-al-2016}, also based on sustained lack of improvement in the smoothed validation loss. 
For further details we refer the reader to Appendix~\ref{app:training_details}.

\subsection{Curriculum Learning Strategy}
\label{subsec:curriculum_learning}

Standard training approaches prove insufficient 
in our case, the main reason being that   
realistic simulated  
GWB often possess amplitudes significantly lower than the intrinsic detector noise, resulting in very low signal-to-noise ratio (SNR). 
Such faint signals 
make it challenging for the network to learn effective separation strategies. To overcome this, we implement a curriculum learning strategy focused on signal amplitude. The core principle is to initially expose the model to signals with artificially elevated amplitudes, 
easier to discern from  
noise. As training progresses over a defined schedule, the amplitude scale factor ($A$) applied multiplicatively to the base signals is gradually reduced, transitioning the training focus towards weaker, more realistic signals. This ``easy-to-hard'' progression allows the model to first grasp the fundamental signal characteristics when they are prominent and subsequently refine its sensitivity, ultimately enhancing convergence and performance, 
in particular important for low-SNR signals.
The strategy defines a dynamic valid amplitude range $[A_{\text{min}}({\cal E}), A_{\text{max}}]$ for each training epoch ${\cal E}$. The term \emph{epoch} refers to a full pass through the entire training dataset, during which the model is exposed once to every available training examples. Monitoring model behavior across epochs provides a natural timescale over which learning progress and generalization can be assessed. In the context of curriculum learning, epochs serve as the control parameter for progressively increasing task difficulty—here, by gradually lowering the minimum signal amplitude presented to the model. This structured pacing allows the network to first identify prominent signal characteristics at high amplitudes and then refine its capacity to detect subtler features under more realistic, low-SNR conditions.

The effective minimum amplitude $A_{\text{min}}({\cal E})$ is determined by first interpolating then clipping. The normalized curriculum progression $\tau({\cal E})$ is defined as
\begin{equation}
    \tau({\cal E}) = \min\left(1, \frac{{\cal E}-1}{\max(1, E_{\text{curr}}-1)}\right)~,
\end{equation}
where $E_{\text{curr}}$ is the number of epochs.
An intermediate log-amplitude is calculated interpolating between $\log_{10}(A_{0})$ and $\log_{10} (A_{m})$

\begin{equation}
\begin{split}
    \mathcal{A}_{\text{interp},10}({\cal E}) ={}& (1 - \tau({\cal E})) \log_{10}(A_{0}) \\
                                     & + \tau({\cal E}) \log_{10}(A_{\rm sampled})~.
\end{split}
\end{equation}
It is then converted to linear scale and clipped to ensure it respects the overall bounds $[A_{\text{min}}, A_{\text{max}}]$, yielding the effective minimum for epoch ${\cal E}$
\begin{equation}
A_{\text{min}}({\cal E}) = \min\left( A_{\text{max}}, \max\left( A_{\rm sampled}, 10^{\mathcal{A}_{\text{interp},10}({\cal E})} \right) \right)~.
\end{equation}
The valid sampling range at epoch $\cal E$ is $[A_{\text{min}}({\cal E}), A_{\text{max}}]$,
within which the amplitude scale factor $A$ is sampled such that 
$\mathcal{A}_{\ln} = \ln (A)$ follows a Normal distribution $\mathcal{N}(\mu_{\ln}({\cal E}), \sigma_{\ln}({\cal E})^2)$. The mean $\mu_{\ln}({\cal E})$ and standard deviation $\sigma_{\ln}({\cal E})$
adapt during the curriculum phase ($1 \le {\cal E} \le E_{\text{curr}}$) based on the progression $\tau({\cal E})$.

Denoting $\mathcal{A}_{\ln, \text{min}}({\cal E}) = \ln (A_{\text{min}}({\cal E}))$ and $\mathcal{A}_{\ln, \text{max}} = \ln (A_{\text{max}})$, 
the dynamic range in natural log space is $\Delta_{\ln}({\cal E}) = \mathcal{A}_{\ln, \text{max}} - \mathcal{A}_{\ln, \text{min}}({\cal E})$. The standard deviation $\sigma_{\ln}({\cal E})$ is set proportionally to this range via a hyperparameter $f_{\sigma}$, with 
minimum value $\epsilon_{\ln, \sigma}$ to prevent collapse
\begin{equation}
\sigma_{\ln}({\cal E}) = \max\left(\epsilon_{\ln, \sigma}, f_{\sigma} ~ \Delta_{\ln}({\cal E})\right)~.
\end{equation}
The mean $\mu_{\ln}({\cal E})$ transitions linearly over the curriculum duration using $\tau({\cal E})$
from a starting position relative to the maximum ($\mathcal{A}_{\ln, \text{max}}$) to an ending position relative to the minimum ($\mathcal{A}_{\ln, \text{min}}({\cal E})$). These 
positions are controlled by hyperparameters $f_{\mu, 0}$ and $f_{\mu, 1}$, representing offsets in units of the current standard deviation $\sigma_{\ln}({\cal E})$
\begin{align}
 \mu_{\ln, \text{start}}({\cal E}) &= \mathcal{A}_{\ln, \text{max}} - f_{\mu, 0}  \sigma_{\ln}({\cal E})  \nonumber \\ 
 \mu_{\ln, \text{end}}({\cal E}) &= \mathcal{A}_{\ln, \text{min}}({\cal E}) + f_{\mu, 1}  \sigma_{\ln}({\cal E})  \\ 
 \mu_{\ln}({\cal E}) &= (1 - \tau({\cal E}))  \mu_{\ln, \text{start}}({\cal E}) + \tau({\cal E})  \mu_{\ln, \text{end}}({\cal E}) \,. \nonumber \label{eq:normal_mu_schedule_ln} 
\end{align}
A log-amplitude value $\mathcal{A}_{\ln, \text{sampled}}$ is drawn from the evolving Normal distribution, $\mathcal{A}_{\ln, \text{sampled}} \sim \mathcal{N}(\mu_{\ln}({\cal E}), \sigma_{\ln}({\cal E})^2)$, and is converted back to linear scale, $A_{\text{sampled}} = \exp(\mathcal{A}_{\ln, \text{sampled}})$. The final amplitude scale factor $A$ applied to the signal component is obtained by restricting $A_{\text{sampled}}$ to the current valid dynamic range $[A_{\text{min}}({\cal E}), A_{\text{max}}]$
\begin{equation}
 A = \max\left(A_{\text{min}}({\cal E}), \min\left(A_{\text{sampled}}, A_{\text{max}}\right)\right)~.
\end{equation}
This strategy facilitates a gradual transition from higher to lower amplitude signals while ensuring sampled amplitudes adhere strictly to the defined bounds.

The amplitude sampling method presented above, is applied when generating scale factors for data 
containing signals. 
This curriculum learning modulates the amplitude scaling for samples belonging to categories $k \in \{0, 1, 2, 3\}$. The selection of a category $k$ for each training sample is based on probabilities $P_k$. The set of probabilities $(P_0, P_1, P_2, P_3)$ is defined from the configuration at the beginning of the training phase and remains fixed throughout its entirety.

\section{Simulated data}
\label{sec: Data Model}

We model data of a network composed of the two advanced LIGO detectors located in Hanford and Livingston, along with Virgo in Cascina, all operating at their designed sensitivity. We consider the A+ design sensitivity curve given in~\cite{T1800042:2018} for the two LIGO detectors.

The simulated data, for both analysis and training, are generated using a simulation code developed for Einstein Telescope \footnote{https://gitlab.et-gw.eu/osb/div10/mdc-generation}. The resulting output includes time series of the detector's measurements, which comprise both instrumental noise and gravitational-wave signals from a population of compact binary coalescences as well as a stochastic GWB from the early Universe. In the next sections we briefly explain the different methods used to simulate noise and signals datasets.

\subsection{Simulation of the Detector Noise}

The noise is modeled as Gaussian and uncorrelated between detectors. For each detector, independent Gaussian frequency series are generated with zero mean and unit variance. They are then shaped according to the power spectral density (PSD) of each detector and transformed back to the time domain via inverse Fourier transform.

For computational efficiency, the dataset is split into  segments of \unit[2048]{s}, with 50\% overlap between successive segments sampled at \unit[512]{Hz}. The overlapping regions are smoothly blended using sine and cosine functions.
This technique helps to eliminate discontinuities and ensures the continuity of the time series~\cite{lalsuite}.

\subsection{Simulation of the Astrophysical Background}

The astrophysical GWB is generated from populations of compact binary coalescences, where the intrinsic parameters, masses, spins, tidal parameters (for neutron stars) and redshifts, are drawn from population synthesis catalogs (see 
\cite{Mapelli:2021syv, Martinovic:2020hru} for binary black holes and \cite{Santoliquido:2020axb} for systems including a neutron star). 
The source extrinsic parameters, right ascension $ra$, declination $\delta$ and inclination angle, are drawn from isotropic distributions while the polarization angle $\psi$ and the initial phase when the signal frequency reaches \unit[5]{Hz}, are uniformally distributed. The time interval between successive coalescences is modeled as a Poisson process, with the time $\tau$ drawn from an exponential distribution $P(\tau) = \exp(-\tau / \lambda)$. The parameter $\lambda$ represents the average time between events. 
Finally, the two signal polarizations $h_+(t)$ and $h_{\times}(t)$ are computed using the IMRPhenomPv2 waveform model with tidal effects NRTidalv2\_V \cite{Dietrich:2019kaq} for binary neutron star (BNS) and the IMRPhenomXPHM \cite{Pratten:2020ceb} model for both BBHs and neutron star - black hole (NSBH). 

The time-domain waveform for each source is added to each detector strain time-series. The response of each detector is calculated using the time-dependent antenna pattern functions $F^j_+(ra, \delta, \psi; t)$ and $F^j_{\times}(ra, \delta, \psi; t)$ for the two polarizations. The total response $h_j(t)$ of the $j$-th detector is
\be
h_j(t) = F^j_+(ra, \delta, \psi; t) h_+(t - t_d^j) + F^j_{\times}(ra, \delta, \psi; t) h_{\times}(t - t_d^j)~,
\ee
where $t_d^j$ is the time delay between the gravitational wave arrival at the detector and the Earth's center. The waveforms from all sources are summed to generate the total gravitational-wave signal for each detector. We refer the reader to \cite{regimbau:2025xxx} for further details on 
the astrophysical GWB data generation.

\subsection{Simulation of the Cosmological Background}

For the cosmological GWB we follow an approach similar to noise generation, starting with signal synthesis in the frequency domain. Unlike instrumental noise, which is usually independent between detectors, the GWB induces a correlation between their outputs that is characterized by the overlap reduction function as 
\begin{equation}
<\tilde{h}_i(f)\tilde{h}_j(f')>= \frac{1}{2} \delta(f-f') \gamma_{ij}(f) S_h(f)
\end{equation}
To simulate the signals of two detectors, we generate two independent complex Gaussian random frequency series \( \tilde{x}_i(f) \) and \( \tilde{x}_j(f) \), each with variance proportional to \( S_h(f) \). The first detector’s signal is
\begin{equation}
\tilde{h}_i(f) = \tilde{x}_i(f)~,
\end{equation}
while the second detector’s signal is constructed as
\begin{equation}
\tilde{h}_j(f) = \gamma_{ij}(f) \tilde{x}_i(f) + \sqrt{1 - \gamma_{ij}^2(f)} \, \tilde{x}_j(f)~.
\end{equation}.
This construction introduces a component that is correlated with the first detector, and a component that remains independent. In doing so, it ensures that the simulated signals reproduce the correct cross-correlation structure implied by the GWB. Mathematically, this is equivalent to performing a Cholesky decomposition of the cross-power spectral density matrix. In the special case where \( \gamma_{ij}(f) = 1 \), corresponding to co-located and co-aligned detectors, the two signals become fully correlated, as expected.

The time-domain strains are then obtained by applying the inverse Fourier transform to these frequency-domain signals.
Consequently, we cross-correlate the generated time-domain strains from each detector to build the final spectra dataset.
For more details and multiple detectors we refer the reader to~\cite{2003CQGra..20S.677B, 2007CQGra..24S.639C}.

\section{Method validation}
\label{sec:validation}

We present the performance of our deep learning framework combined with parameter estimation techniques to infer the GWB components. The core of the validation involves assessing the ability of the trained MSMHAutoencoder to separate GWB signals from detector noise, and subsequently estimating the parameters using Bayesian inference.
We restrict our analysis to the BBH component of the astrophysical GWB for the sake of simplicity, without impacting the methodological results. To reduce the computational cost associated with training, we adopt an average time interval between successive events of \unit[25]{s}, which corresponds to the higher bound of the expected BBH merger rate. This choice limits the duration of the time series required to obtain a representative GWB while still capturing the key statistical properties of the BBH signal.

The MSMHAutoencoder is trained using the methodology described in Section \ref{subsec:training_methodology}, including the physics-informed loss function and the curriculum learning strategy. The key hyperparameters defining the specific MSMHAutoencoder architecture and the training configuration are summarized in Table~\ref{tab:hyperparameters}. These hyperparameters are not obtained through systematic tuning methods such as grid search or Bayesian optimization because of the computational cost of such methods, given the large number of tunable parameters. Instead, hyperparameters are selected empirically through iterative experimentation, with manual adjustments guided by observed training behavior, such as convergence speed, generalization performance and signal reconstruction quality. For the development and evaluation of the model, we used the following data split: a total of \unit[$47.4$]{days} of data was initially set aside. From this, 80\% (\unit[$37.9$]{days}) was used as the training set, which is the data the model directly learns from to adjust its internal parameters. The remaining 20\% (\unit[$9.5$]{days}) was designated as the validation set. 
The performance results are generated using a separate testing dataset of \unit[$23.7$]{days}. This dataset is held out during all model development and hyperparameter tuning to ensure an unbiased estimation of the model's generalization capabilities.

\begin{figure}[htbp]
    \centering
    \includegraphics[width=\linewidth]{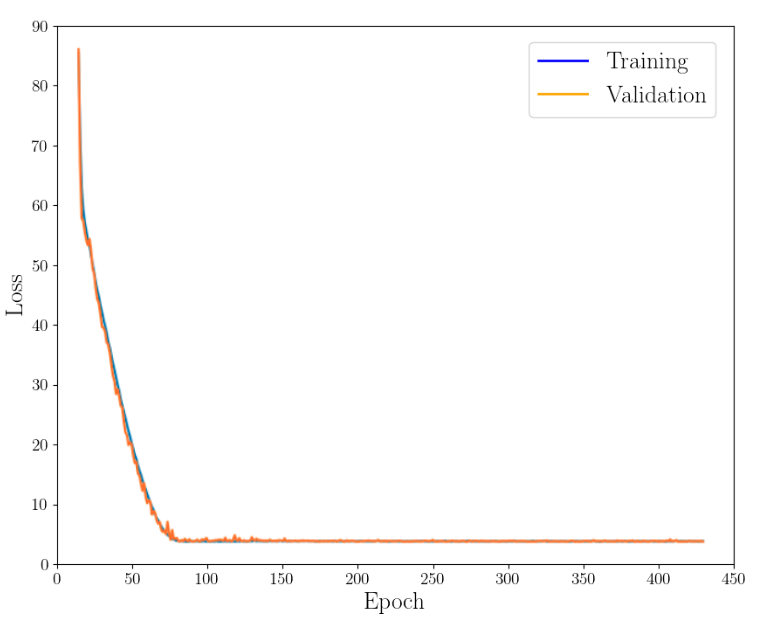}
    \caption{Evolution of the training loss (blue) and validation loss (orange) during the training of the MSMHAutoencoder model performed on \unit[47.4]{days} of simulated data, comprising a BBH and a cosmological GWB mixed with LIGO Livingston and LIGO Hanford simulated detector noise. The stabilization of the validation loss indicates successful learning without significant overfitting. 
    }
    \label{fig:loss_curve}
\end{figure}

The loss functions for training and validation converge smoothly and similarly as shown in Figure~\ref{fig:loss_curve}. 
Crucially, the validation loss function closely tracks the training loss function, reaching a plateau at a similarly low value without significant divergence.
This convergence behavior provides evidence for the adequacy for learning of the training dataset, the chosen model architecture, optimization strategy, and the empirically selected hyperparameters.

\begin{table}[htbp] 
    \centering
    \caption{Hyperparameters of the MSMHAutoencoder model and their training configuration or initial values.}
    \label{tab:hyperparameters}
    \begin{tabular}{lc}
        \toprule 
        \\
        Parameter                       & Value \\
        \\
        \hline
        \\
        \midrule 
        \multicolumn{2}{l}{\textit{Architecture}} \\
        \\
        Input spectra per prediction ($M$) & 12 \\
        Frequency bins ($N$)            & 1005 \\ 
        Encoder depth ($n_{\rm layers}$) & 3 \\ 
        Base channels ($c_1$) & 64 \\
        Inception kernel sizes          & (3, 5, 7) \\ 
        \midrule 
        \\
         \hline
         \\
        \multicolumn{2}{l}{\textit{Training}} \\
        \\
        Optimizer                       & Adam \\
        Initial learning rate ($\eta_0$) & $5 \times 10^{-5}$ \\
        Adam betas ($\beta_1, \beta_2$) & (0.9, 0.999) \\ 
        Weight decay                    & $5 \times 10^{-6}$ \\
        Gradient clip norm (L2)         & 0.1 \\
        Batch size ($B$)                & 32 \\
        Max training epochs             & 2000 \\
        LR scheduler patience (patience)& 5 \\
        LR scheduler factor ($\rm lr_{factor}$) & 0.8 \\
        Min learning rate ($\rm min_{lr}$)    & $10^{-9}$ \\
        Early stopping patience         & 30 \\
        Early stopping min delta        & $10^{-5}$ \\
        Loss weights ($\lambda_l, \lambda_c, \lambda_s$) & (0.1, 0.1, 0.5) \\ 
        Scenario 0 signal factor        & 0.5 \\ 
        \\
        \midrule 
         \hline
         \\
        \multicolumn{2}{l}{\textit{Curriculum learning}} \\
        \\
        Duration ($E_{\text{curr}}$)    & 100 \\ 
        Start min amplitude ($A_0$)     & $ 10^{2}$ \\ 
        Final min amplitude ($A_m$)     & $ 10^{-2}$ \\ 
        Max amplitude ($A_{\text{max}}$)  & $10^{2}$ \\ 
        Sigma Factor ($f_{\sigma}$)     & 0.35 \\ 
        Mu Start Factor ($f_{\mu, 0}$)  & 1.0 \\  
        Mu End Factor ($f_{\mu, 1}$)    & 0.5 \\  
        Min $\sigma_{\ln}$ ($\epsilon_{\sigma}$) & $10^{-6}$ \\ 
        \bottomrule 
    \end{tabular}
\end{table}

Parameter estimation on the network's output \( \hat{s}(f) \) is conducted using Bayesian inference. The posterior probability distribution \( p(\theta | \hat{s}) \) for the signal model parameters \( \theta \) is explored using Markov Chain Monte Carlo (MCMC) methods. Specifically, we employ an affine-invariant ensemble sampler, as implemented in the \texttt{emcee} package~\cite{emcee},
using 50 walkers, each taking 2000 steps after discarding an initial 1000 steps as burn-in. According to Bayes' theorem, the posterior is proportional to the product of the likelihood \( p(\hat{s} | \theta) \) and the prior \( p(\theta) \) 
\[
p(\theta | \hat{s}) \propto p(\hat{s} | \theta) \, p(\theta)~.  
\]
The total energy density spectrum $\Omega_{\rm GW}(f)$ is modeled as
\begin{equation}
    \Omega_{\rm model}(f | \theta) = \Omega_{\rm BBH}(f | \Omega_{\alpha}, \alpha) + \Omega_{\rm Cosmo}(f | \Omega_0,\alpha=0)~,   
\end{equation}
where $\Omega_{\rm Cosmo}(f | \Omega_0) = \Omega_0$. The BBH component is modeled as a power law, $\Omega_{\rm BBH}(f | \Omega_{\rm \alpha}, \alpha) = \Omega_{\rm \alpha} ( f / f_{\rm ref} )^{\alpha}$, with $f_{\rm ref} = 25$ Hz.
In the following, we initially consider that the BBH spectral index is fixed at its theoretically expected value for compact binary coalescences, $\alpha = 2/3$. 

The combined model spectrum, \( \Omega_{\mathrm{model}}(f | \theta) \), is transformed into the logarithmic domain
\begin{multline}
    \log_{10} (\Omega_{\mathrm{model}}(f|\theta)) = \\ \log_{10} ( 10^{\log_{10}(\Omega_0)}+ 10^{\log_{10}(\Omega_{\rm \alpha})} \left( \frac{f}{f_{\mathrm{ref}}} \right)^{2/3} + \epsilon )~.
\end{multline}

Assuming Gaussian uncertainties $\sigma_{\log}(f)$ on the MSMHAutoencoder's predicted $\log_{10}(\hat{s}(f))$, which are taken to be independent across frequencies, the natural log-likelihood $\mathcal{L} = \ln(p(\hat{s} | \theta))$ is given by
\begin{multline}
    \mathcal{L}(\theta) = -\frac{1}{2} \sum_{f} \left[ \frac{(\log_{10}(\hat{s}(f)) - \log_{10}(\Omega_{ \rm model}(f | \theta)))^2}{\sigma_{\log}^2(f)} \right. \\ \left. + \ln(2\pi \sigma_{\log}^2(f)) \right] \, .
\end{multline}
The uncertainty $\sigma_{\log}(f)$ is estimated from the variability of its predictions over 996 distinct sequences of $M=12$ input spectra.

For the priors $p(\theta)$, we employ uniform distributions for the logarithmic amplitudes within plausible ranges and enforce $\Omega_0 < \Omega_{\rm \alpha}$
\begin{equation}
\log_{10} (p(\theta)) =
\begin{cases}
0 & \text{if } -13 \le \log_{10} (\Omega_0) \le -9 \\
& \text{\& } -11 \le \log_{10} (\Omega_{\rm \alpha}) \le -7 \\
& \text{\&} \, \log_{10} (\Omega_0) < \log_{10} (\Omega_{\rm \alpha}) \\
-\infty & \text{otherwise}~.
\end{cases}
\end{equation}
To quantify statistical evidence for detecting signal components against noise, we employ a Bayesian model comparison. This involves computing the Bayesian evidence, $\mathcal{Z}$, for competing hypotheses, using nested sampling algorithms such as those implemented in \texttt{dynesty}~\cite{dynesty}. 
The preference for the signal model over the noise model is quantified by the Bayes factor, BF = $\mathcal{Z}_1 / \mathcal{Z}_0$, where $\mathcal{Z}_1$ represents the evidence calculated for the signal model for a given dataset and $\mathcal{Z}_0$ corresponds to the noise model. 

For interpretation aligned with common practice (e.g., the Jeffreys scale \cite{jeffreys1998theory}) we consider $\log_{10}$(BF).
It allows us to establish detection thresholds by identifying the minimum signal amplitude required to achieve a $\log_{10} $(BF) value corresponding to desired levels of evidence such as $\log_{10}(\rm BF) > 1$ for positive, and $> 3$ for decisive evidence.

To rigorously assess the method's sensitivity, we 
do injection studies that involve generating datasets containing known 
GWB signals, comprising both BBH and cosmological components, into simulated LIGO Hanford and Livingston detectors' noise. 
The amplitudes of the signal components vary
to evaluate the performance under different signal strength conditions. 

From the MCMC samples of each parameter's posterior distribution, we determine the best-fit estimate as the median. Uncertainties are reported as the 68\% credible interval, with its bounds defined by the 16th and 84th percentiles of the MCMC samples, reflecting the potentially asymmetric nature of the posterior. The MCMC inference is restricted to the \unit[$(10-100)$]{Hz} frequency range.

\begin{figure}[htbp]
    \centering
    \includegraphics[width=0.48\textwidth]{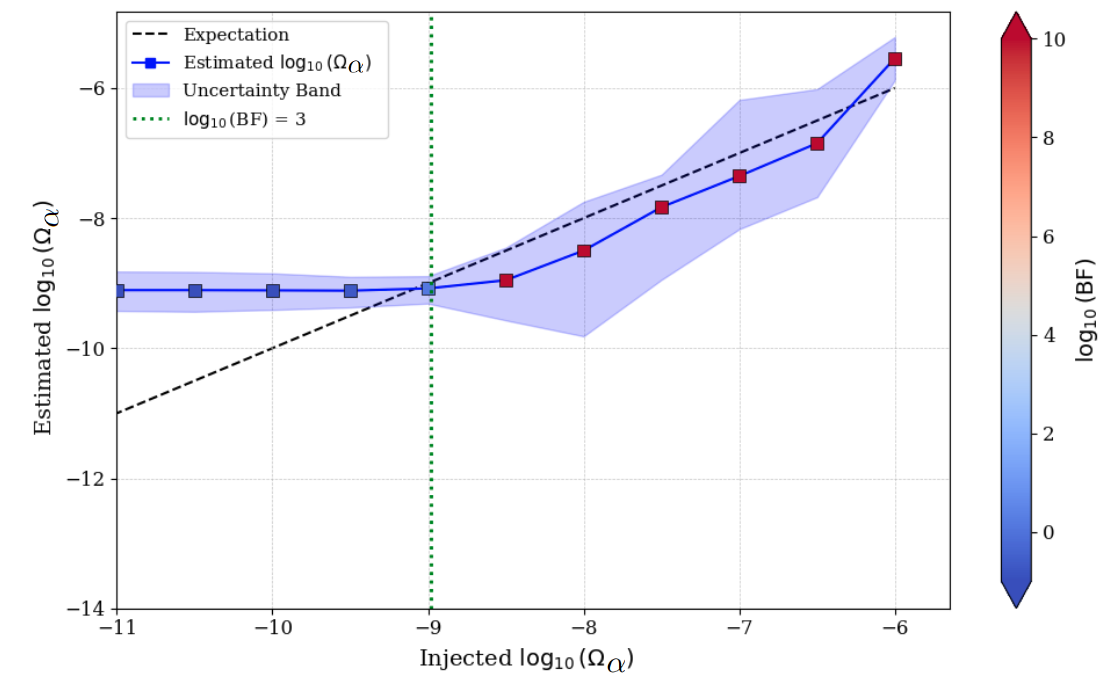} 
    \caption{Recovered median $\log_{10}(\Omega_{\rm \alpha})$ (colored markers) with $1\sigma$ uncertainty (blue area) as a function of the injected signal amplitude $\log_{10}(\Omega_{\rm \alpha})$. The $\log_{10}$(BF) of each estimate is displayed using a color map. The black dashed line represents the expected 
    prediction. The vertical green dashed line indicates the $\log_{10}(\Omega_{\rm \alpha})$ 
    corresponding to $\log_{10} $(BF) = 3.}
    \label{fig:sens_BBHonly}
\end{figure}

We first consider the case of a single BBH component whose amplitude varies. Figure~\ref{fig:sens_BBHonly} shows the estimated $\log_{10} (\Omega_{\rm \alpha})$ that matches well the injected signal amplitude, with large BF values, until $\Omega_{\rm \alpha}$ becomes smaller than $10^{-9}$ for which $\log_{10} $(BF) gets smaller than $3$ considering \unit[23.7]{days} of data. For lower injected amplitude, the estimated value remains constant with a small $1\sigma$ uncertainty interval. For signal amplitude $\Omega_{\rm \alpha} < 10^{-9}$ the MSMHAutoencoder is no longer sensitive to the faint GWB signal. In this regime, the MCMC is effectively fitting residual noise patterns that have been shaped by the MSMHAutoencoder's signal reconstruction path. This process can yield a statistically precise fit to these noise features, resulting in a narrow credible interval, while the low $\log_{10} $(BF) correctly reflects the absence of significant evidence for an actual GWB signal.

\begin{figure}[htbp]
    \centering
    \includegraphics[width=0.48\textwidth]{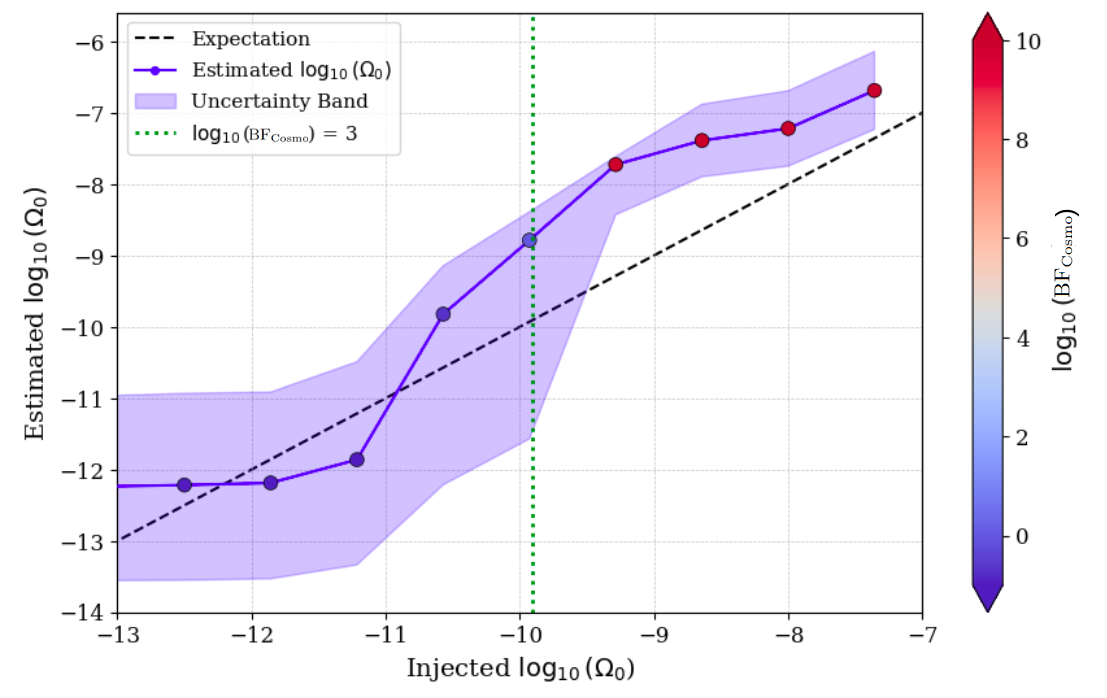} 
    \caption{
    Recovered median $\log_{10}(\Omega_0)$ (colored markers) with $1\sigma$ uncertainty (blue area) as a function of the injected signal $\log_{10}(\Omega_0)$. The $\log_{10} (\rm BF_{Cosmo})$ of each estimate is shown using a color map. The black dashed line represents the expected prediction. The vertical green dashed line indicates the $\log_{10}(\Omega_{0})$ value corresponding to $\log_{10}(\rm BF_{Cosmo}) = 3$.
    }
    \label{fig:sens_BBHCosmo}
\end{figure}

Let us now consider a dataset containing a BBH GWB with a fixed amplitude $\Omega_{\rm \alpha} = 10^{-9}$ and a cosmological GWB whose amplitude varies. The estimated $\log_{10} (\Omega_{\text{0}})$ as a function of the injected amplitude is shown in Figure~\ref{fig:sens_BBHCosmo}. 
To assess whether a cosmological component is supported by the data in addition to the BBH background, we perform a second model comparison. Specifically, we compute a Bayes factor defined as the ratio of the Bayesian evidences $\mathcal{Z}_1$ and $\mathcal{Z}_0$, where $\mathcal{Z}_1$ corresponds to the model including both BBH and cosmological GWB components, and $\mathcal{Z}_0$ to the BBH-only model. This Bayes factor, $\rm BF_{Cosmo}$, quantifies the preference for a cosmological contribution beyond what can be explained by the BBH background alone.

Using the same Jeffreys scale interpretation, we consider $\log_{10}(\rm BF_{Cosmo}) > 3$ as strong evidence in favor of the BBH+Cosmo model.
Under this criterion, we find that cosmological GWB amplitudes as low as $\Omega_0 \approx 1.3 \times 10^{-10}$ can be distinguished from a pure BBH signal using \unit[23.7]{days} of data.

At such signal strengths, the $1\sigma$ credible interval for the estimated $\log_{10} (\Omega_{\text{0}})$ is substantial, spanning approximately two orders of magnitude. At such faint injected amplitudes, the MSMHAutoencoder is unable to identify the cosmological GWB, one order of magnitude lower than the BBH GWB, and much weaker than the detector noise.
During the subsequent MCMC analysis, attempting to constrain the parameters of the cosmological GWB in the presence of comparatively stronger BBH GWB foreground and noise, results in a very flat likelihood function for $\log_{10} (\Omega_{\text{0}})$.

\begin{figure}[htbp]
    \centering
    \includegraphics[width=0.45\textwidth]{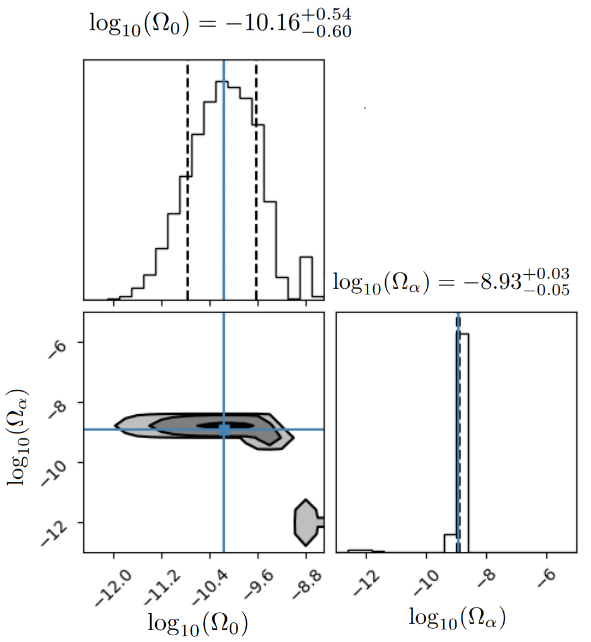} 
    \includegraphics[width=0.45\textwidth]{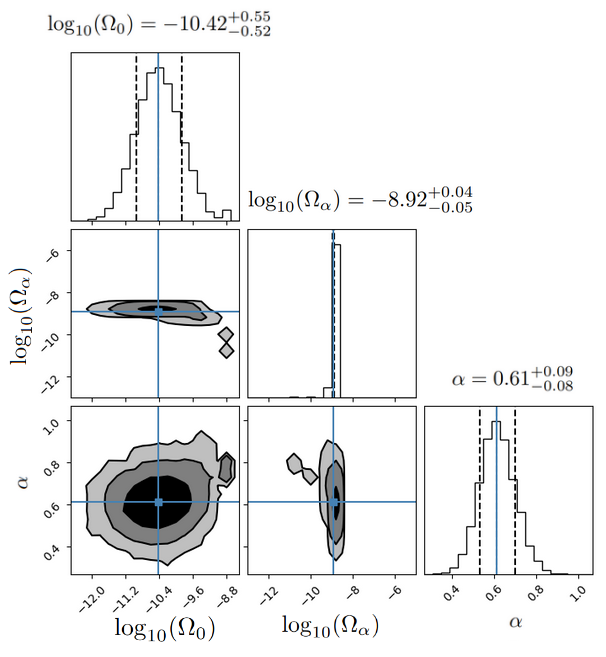} 
    \caption{Corner plots showing the marginalized posterior probability distributions for the estimated GWB components' parameters for a dataset containing a cosmological and a BBH GWB with amplitude $\Omega_0 = 10^{-10}$ and $\Omega_{\rm \alpha} = 10^{-9}$, respectively. Top: the spectral index $\alpha$ is fixed at $2/3$. Bottom: $\alpha$ is estimated.
    }
    \label{fig:corner_plot_example}
\end{figure}

Let us then relax the assumption $\alpha=2/3$ for the BBH GWB adopting a uniform prior for the spectral index $\alpha$. In Figure~\ref{fig:corner_plot_example} we compare the posterior probability distributions for the parameters $\Omega_0$, $\Omega_{\rm \alpha}$ and $\alpha$ for a dataset that contains a cosmological and a BBH GWB with amplitudes $\Omega_0 = 10^{-10}$ and $\Omega_{\rm \alpha} = 10^{-9}$, respectively. When $\alpha$ is treated as a free parameter, the median of the estimated cosmological GWB amplitude, $\Omega_0$, changes from $(6.9^{+17}_{-5.2}) \times 10^{-11}$ for fixed $\alpha$, to $(3.8^{+9.7}_{-2.7}) \times 10^{-11}$. Despite the shift in the median value, the estimate is consistent with the one obtained for fixed-$\alpha$ within  \( 1\sigma\) uncertainty.
Furthermore, considering $\alpha$ as a free parameter, its posterior distribution is well-constrained around the expected value $2/3$. The statistical significance for the presence of the GWB is not impacted by allowing $\alpha$ to vary ($\log_{10}(\text{BF}) = 3$ in both cases).

\begin{figure}[htbp]
    \centering
    \includegraphics[width=0.5\textwidth]{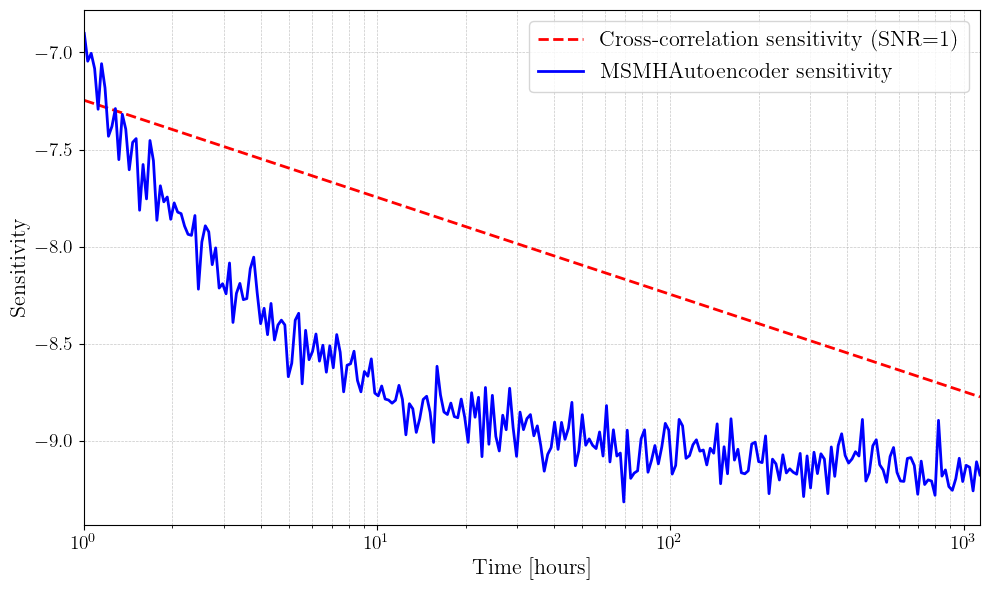} 
    \caption{Sensitivity as function of the amount of unique training data used (expressed in equivalent of observation time). The red dashed line represents the cross-correlation sensitivity.}
    \label{fig:sensitivity_vs_training_time}
\end{figure}

To quantitatively assess the amount of training data necessary to reach a given detection sensitivity, we study how the model's sensitivity at $f_{\text{ref}}=25\,$Hz trend evolves as it is trained with larger volumes of simulated data. At each epoch of training, the MSMHAutoencoder is tested 
on new spectra of simulated data. Each segment corresponds to \unit[2048]{s} of data, such that the total cumulative exposure grows with training. At regular checkpoints during training, we freeze the MSMHAutoencoder and evaluate its performance on the validation set at varying amplitudes. 
For this analysis, we define a sensitivity metric based on reconstruction fidelity. We determine the minimum log amplitude of an injected BBH GWB signal for which the Mean Squared Error (MSE) between the model's predicted spectrum and the true injected spectrum falls below a threshold of $0.01$
\begin{equation}
   \text{MSE}_{\hat{s}}=\frac{1}{N} \sum_{j=1}^{N} \left[\hat{s}(f_j | \Omega_{\rm \alpha}) - s(f_j | \Omega_{\rm \alpha}) \right]^2 < 0.01~.
\end{equation}
This MSE threshold is arbitrarily chosen and represents a difference of $\sim 3\%$ between the target and the predicted signal. Each sensitivity estimate is associated with the total cumulative volume of distinct training data the model uses up to that checkpoint. The resulting MSMHAutoencoder sensitivity evolution as a function of training time $T$ is depicted (solid blue curve) in Figure~\ref{fig:sensitivity_vs_training_time} and is compared to the evolution of the cross-correlation sensitivity as function of observing time for a signal detected with SNR = 1 (red dashed curve), 
which is proportional to $1/\sqrt{T_{\text{obs}}}$ where $T_{\text{obs}}$ is the observing time~\cite{1999PhRvD..59j2001A, 2013PhRvD..88l4032T, 2022Galax..10...34R}.
Figure~\ref{fig:sensitivity_vs_training_time} indicates that our deep learning approach achieves better sensitivity quicker than the cross-correlation method, particularly at lower cumulative observation times (before $\sim 10$ hours). Initially, the MSMHAutoencoder sensitivity is proportional to \(T^{-1.6}\). As more training data is used (after $\sim 10$ hours), the sensitivity gain of the MSMHAutoencoder slows down to reach \(T^{-0.2}\). More precisely, the sensitivity improves from $\Omega_{\rm \alpha} \approx 10^{-7}$ to $\Omega_{\rm \alpha} \approx 10^{-9}$ in less than $100$ hours of training time. To reach $\Omega_{\rm \alpha} \approx 5 \times 10^{-10}$, more than $10^3$ hours are needed.
This marked slowdown indicates a stage beyond which further substantial increases in training data volume provide only marginal gains in sensitivity for the current model configuration.

\section{Discussion and Summary}
\label{sec: Conclusions}
The dominant GWB due to the superposition of gravitational waves emitted by compact binary coalescences and mergers,
is expected to be accompanied with other weaker GWB components. This is especially the case of a GWB due to early Universe phenomena. 
In this study,
we show that our MSMHAutoencoder architecture is able to estimate the contribution of two different components of the GWB in the data of the LIGO-Virgo-KAGRA network of detectors when they reach their design sensitivity circa 2030. More precisely, we show that we can detect, with high confidence, the astrophysical background due to BBH mergers if its amplitude is as low as $\Omega_{\rm \alpha} \sim 10^{-9}$ at \unit[25]{Hz}. This result can be extrapolated to other CBC sources, BNS and NSBH, as long as their energy density spectrum follows a power law with index $\sim 2/3$. This sensitivity corresponds to the expected upper range amplitude of the BNS+BBH GWB component assuming the CBC rate measured by the LVK Collaboration~\cite{KAGRA:2021kbb}. Our sensitivity results can also be compared to the LVK cross-correlation search sensitivity estimate. In \cite{KAGRA:2021kbb}, the LVK Collaboration reports a $2\sigma$ significance sensitivity of $5\times 10^{-10}$ for the same A+ detectors' sensitivity. Our search sensitivity is two times higher but in our study it corresponds to a confident detection ($\log_{10}(\rm BF)=3$), that would be close to a $5\sigma$ significance sensitivity. In this case, the cross-correlation method sensitivity is  $1.3\times 10^{-9}$~\cite{2013PhRvD..88l4032T,Thrane2013} for \unit[1]{year} of data. To achieve our MSMHAutoencoder sensitivity $\Omega_{\rm \alpha} \sim 10^{-9}$, one would need \unit[1.6]{years} of coincident data from both detectors.
Moreover, our MSMHAutoencoder architecture is able to estimate simultaneously a weaker GWB component, of cosmological origin. We show that we are sensitive, with high confidence, to measure a CBC GWB of amplitude $10^{-9}$ and a cosmological GWB of amplitude $1.3 \times 10^{-10}$. 
In addition, we show that the constraint on the spectral index of the expected CBC energy density spectrum can be relaxed without impacting the accuracy of the CBC and cosmological GWB measurement.
We also show that such an autoencoder-based algorithm allows us to reach a good accuracy of the GWB measurement much quicker than the cross-correlation method.

Those results, obtained with a training dataset of \unit[47.4]{days}, may not be yet enough to detect any of the cosmological GWB components in future LVK observing runs. They remain though interesting as many improvements are still possible.
Indeed, in this proof-of-concept study, we have not exhaustively explored the full potential of deep learning for GWB detection and its characterization. The MSMHAutoencoder architecture, while demonstrating promising results, represents an initial design that was shown to be effective rather than the outcome of an extensive optimization process. Similarly, the physics-informed loss function could be subject to further refinements. Hyperparameters optimization, governing both the model architecture and the training dynamics was beyond the scope of this initial investigation due to the significant computational resources required for such large-scale tuning. Consequently, it is highly probable that alternative configurations exist that could yield better performances. Furthermore, the exploration of model scalability was constrained by available computational capabilities, particularly GPU memory and processing power. Beyond these architectural and hyperparameter refinements, several avenues for future development remain. Larger and more complex neural network architectures, potentially incorporating a greater number of parameters, more sophisticated building blocks or nested MSMHAutoencoder for individual GWB components separation could be explored. The current study relied on specific simulations of astrophysical  and cosmological backgrounds  with idealized detector noise. Extending the framework to incorporate more diverse and realistic astrophysical population models, a broader range of cosmological GWB scenarios, such as those arising from first-order phase transitions  or cosmic strings , and more complex, non-Gaussian or correlated noise characteristics will be crucial for assessing the robustness and real-world applicability of our approach. The adaptability of the MSMHAutoencoder to different types of GWB signals also warrants further investigation. Finally, while our method shows competitive sensitivity gains, particularly at shorter observation times compared to traditional cross-correlation techniques, continued benchmarking against established methods and other emerging machine learning approaches across varied datasets will be essential to fully delineate its strengths and limitations. We also need to test the MSMHAutoencoder method on more realistic data set. This includes using more complete models of astrophysical GWB sources as well as real data with which we could test the impact of the artifacts present in real LVK detectors' data on the sensitivity of our autoencoder method. Such studies will be the subject of a forthcoming article.\\

\sectionname{Acknowledgments}
We would like to thank our colleagues  Guillaume Boileau, Nelson Christensen and Jishnu Suresh for their constant and useful insights throughout the projet. We also thank our LVK colleagues, Chris Messenger and Alberto Mariotti, for reading carefully the manuscript and providing useful feedback. The authors are grateful for computational resources provided by the LIGO Laboratory and supported by the National Science Foundation Grants PHY-0757058 and PHY-0823459. MS acknowledges support from the
Science and Technology Facility Council (STFC), UK, under
the research grant ST/X000753/1. This work was supported by the French government through the France 2030 investment plan managed by the National Research Agency (ANR), as part of the Initiative of Excellence of Université Côte d’Azur under reference number ANR-15-IDEX-01.
This material is based upon work supported by NSF’s LIGO Laboratory which is a major facility fully funded by the National Science Foundation.

This manuscript was assigned LIGO-Document number LIGO-2500339.

\appendix

\section{Detailed Network Architecture Implementation}
\label{app:architecture_details}

In the following 
we present in some detail the core operations underlying the architecture introduced in Section~\ref{subsec:model_arch}, detailing the mathematical structure of each component and explicitly defining all quantities used. 

The architecture constructs latent representations at multiple scales, denoted by the index \(\ell \in \{1, \dots, L= n_{\rm nlayers}\}\), corresponding to successive levels of abstraction. That is, each level \(\ell\) processes a representation of the input that has a coarser frequency resolution but a broader effective receptive field, allowing it to encode patterns ranging from fine spectral details at shallow levels to coarse global structures at deeper levels. Each scale \(\ell\) operates on a compressed version of the input at reduced spectral resolution \(n_\ell\), with \(n_1 = N\) and \(n_{\ell+1} = \lfloor n_\ell / 2 \rfloor\) due to successive downsampling operations. 

At each scale, the number of latent features is represented by the number of channels \(c_\ell\), such that \(c_1 = 64\) at the bottleneck, which increases with depth to allow the model to capture a richer and more diverse set of learned patterns. These channels correspond to learned filters that act as trainable frequency-domain detectors: each channel can be interpreted as projecting the signal onto a particular learned spectral structure, analogous to decomposing the input on a learned basis. The number of channels doubles at each level such that \(c_{\ell+1} = 2c_\ell\), balancing loss of spectral resolution with gain in feature capacity. 

The frequency axis \(n_\ell\) represents the resolution along the spectral domain, while the channel index \(c_\ell\) spans across filters detecting specific patterns. The resulting latent representation at each scale is therefore a tensor \(z^{\ell} \in \mathbb{R}^{c_\ell \times n_\ell}\), which captures information corresponding simultaneously to a specific frequency scale set by \(n_\ell\) and to a semantic abstraction depth set by \(\ell\). This structure enables the network to detect signal features ranging from narrowband spectral lines to broad stochastic patterns, while progressively disentangling them from non-stationary noise contributions.

\paragraph{Encoders.}
The input tensor \(x \in \mathbb{R}^{M \times N}\) consists of \(M\) consecutive power spectra, each with \(N\) frequency bins. It is processed by the encoder \(\rm Encoder_{\text{total}}\), which constructs a hierarchy of latent representations \(\tilde{z}_{\text{total}} = \{ z^{\ell} \}_{\ell=1}^{L}\), where each \(z^{\ell} \in \mathbb{R}^{c_\ell \times n_\ell}\) is a tensor representing \(c_\ell\) abstract feature channels at a compressed frequency resolution \(n_\ell\).

At the first scale \((\ell = 1)\), a multi-scale feature extractor is applied
\begin{equation}
  z^{1} = \mathrm{Inc}(x)~, \quad c_1 = C_{\text{base}},\ n_1 = N~.
\end{equation}
The operator \(\mathrm{Inc}()\) is an Inception block~\cite{Szegedy_2015_CVPR} designed to extract features at different spectral resolutions using parallel 1D convolutional branches, with kernel sizes \((3,5,7)\). 
Each branch applies a 1D convolution of length \(k\). The output of the convolution at position \(i\) and output channel \(c\) is given by
\begin{equation}
  \mathrm{Inc}_k(x)^c_i = \sum_{d=1}^{C_{\text{in}}} \sum_{m=0}^{k-1} W_k^c[d, m] \cdot x^d_{i + m - \lfloor k/2 \rfloor} + b_k^c~,
\end{equation}
where \(W_k\) are the learnable convolutional weight tensors and \(b_k\) are the corresponding biase tensors, \(c_{\text{in}}\) is the number of input channels. Both are of shape \((C_{\mathrm{in}}, k)\), where \(C_{\mathrm{in}}\) is the number of input channels, \(C_{\mathrm{mid}}\) is the number of output channels and \(k\) is the kernel size.

The outputs from all branches are concatenated along the channel axis, resulting in \(|\mathcal{K}| \times c_{\text{mid}}\) channels, then projected back to \(c_1\) via a \(1 \times 1\) convolution, denoted \(\mathrm{Conv1D}_1\)~\cite{Paszke_2019_NeurIPS}. These output channels serve as learned feature detectors operating over different receptive fields, capable of capturing local and mid-scale spectral structures.

For deeper levels \(\ell > 1\), the latent tensors \(z^{\ell}\) are computed recursively by applying another Inception block followed by a downsampling operator
\begin{equation}
  z^{\ell+1} = \mathrm{Down}\left( \mathrm{Inc}(z^{\ell})\right)~.
\end{equation}
The downsampling operator consists of a max pooling operation followed by a \(1 \times 1\) convolution. Max pooling is defined as
\begin{equation}
  z_{\text{pool}}^{c}[i] = \max(z^{c}[2i], z^{c}[2i+1])~,
\end{equation}
which reduces the resolution by a factor of two by selecting the maximum activation over non-overlapping pairs of adjacent bins along the frequency axis. This is followed by a \(1 \times 1\) convolution that doubles the number of channels
\begin{equation}
  z^{\ell+1} \in \mathbb{R}^{c_{\ell+1} \times n_{\ell+1}}~ \quad \text{with } \ c_{\ell+1} = 2c_\ell,\ n_{\ell+1} = \lfloor n_\ell / 2 \rfloor~.
\end{equation}
Each successive scale \(\ell\) thus captures coarser frequency representations with higher abstraction, as increasing \(c_\ell\) allows the network to express more complex or abstract features. The complete encoder output is
\begin{equation}
  \tilde{z}_{\text{total}} = \{ z^{\ell} \}_{\ell=1}^{L}~, \quad S_{\text{total}} = [s^{1}, \dots, s^{L}]~,
\end{equation}
where each \(s^{\ell} = z^{\ell}\) is retained as a \emph{skip connection}~\cite{Ronneberger_2015_MICCAI}, a direct link that feeds encoder outputs to the decoder at the same scale. Skip connections help preserve localized frequency information that might be degraded during downsampling, allowing the decoder to reconstruct fine-grained spectral details more effectively.

\paragraph{Decoders.}
To reconstruct the signal spectrum \(\hat{s} \in \mathbb{R}^N\), the decoder traverses the corrected latent hierarchy \(\tilde{z}_{\text{signal}} = \{ z^{\ell}_{\text{signal}} \}_{\ell=L}^{1}\), from coarsest (lowest resolution, deepest scale) to finest scale. The decoding process is initialized as
\begin{equation}
  z_{\text{dec}}^{L} = z_{\text{signal}}^{L}~.
\end{equation}
At each scale \(\ell = L-1, L-2, \dots, 1\), the decoder upsamples the previous output and merges it with the corresponding skip feature
\begin{align}
  z_{\text{up}}^{\ell+1} &:= \mathrm{Up}(z_{\text{dec}}^{\ell+1}) \in \mathbb{R}^{c_\ell \times 2n_\ell}~, \nonumber\\
  z_{\text{up}}^{\ell+1} &\leftarrow \mathrm{Interp}(z_{\text{up}}^{\ell+1}, n_\ell) \quad \text{if needed}~, \nonumber\\
  z_{\text{merge}}^{\ell} &:= \mathrm{MergeBlock}(\mathrm{Concat}(z_{\text{signal}}^{\ell}, z_{\text{up}}^{\ell+1}, \hat{n}))~, \nonumber\\
  z_{\text{dec}}^{\ell} &:= z_{\text{merge}}^{\ell}~.
\end{align}
The upsampling operator \(\mathrm{Up}()\) is implemented as a 1D transposed convolution ~\cite{Zeiler_2014_ECCV} with kernel size 2 and stride 2. Here, the kernel size defines the width of the convolutional filter, determining how many neighboring frequency bins are combined in each operation. The stride specifies the shift applied to the filter across the input; a stride of 2 doubles the frequency resolution by spacing out the convolution outputs. If the result \(z_{\text{up}}^{(\ell+1)}\) does not match the resolution \(n_\ell\), it is resized using linear interpolation ~\cite{Parker_1983_TMI}
\begin{equation}
  \mathrm{Interp}(z, n_{\text{target}}): \mathbb{R}^{C \times n} \rightarrow \mathbb{R}^{C \times n_{\text{target}}}~.
\end{equation}
The MergeBlock operator merges the upsampled latent with the skip feature, and optionally the predicted noise \(\hat{n}\). It performs two sequential convolutions with batch normalization~\cite{Szegedy_2015_CVPR} \texttt{BatchNorm1d} and  rectified linear unit activations function, or \texttt{ReLU} ~\cite{Glorot_2011_AISTATS}
\begin{multline}
  \mathrm{MergeBlock}(z)=\\
  \mathrm{Conv1D}_3 \circ \mathrm{BatchNorm1d} \circ \mathrm{ReLU} \circ \mathrm{Conv1D}_3(z)~,
\end{multline}

where \(\mathrm{Conv1D}_3\) denotes a 1D convolution of kernel size 3 and the operator $\circ$ denotes function composition. The merged representation \(z_{\text{merge}}^{\ell}\) has the shape \(\mathbb{R}^{c_\ell \times n_\ell}\).

Once the finest scale is reached, the final latent tensor \(z_{\text{dec}}^{(1)}\) is mapped to a single-channel log-spectrum using a projection head
\begin{equation}
  \hat{s}[i] = [\mathrm{Conv1D}_1 \circ \mathrm{ReLU} \circ \mathrm{Conv1D}_3](z_{\text{dec}}^{1}) \in \mathbb{R}^N.
\end{equation}
This step decodes the full-resolution signal spectrum \(\hat{s}\) from the multi-scale latent features. Optionally, the sequence of intermediate decoded features can be retained for inspection or regularization
\begin{equation}
  \tilde{z}_{\text{signal}} = \{ z_{\text{dec}}^{\ell} \}_{\ell=L}^{1}~.
\end{equation}
Each \(z_{\text{dec}}^{\ell}\) lies in \(\mathbb{R}^{c_\ell \times n_\ell}\), progressively reconstructing the spectral content from coarse to fine resolution. This hierarchical decoding allows the model to integrate contextual information from broad spectral patterns while preserving the localized details necessary for accurate GWB reconstruction.

\paragraph{Corrector.}
To improve the separation of the faint signal components, we introduce a correction mechanism that operates independently at each latent scale \(\ell\). This mechanism refines the naive latent subtraction \(z^{\ell}_{\text{total}} - z^{\ell}_{\text{noise}}\) by learning a residual correction from the joint feature space.

At each scale, the total and noise latents \(z^{\ell}_{\text{total}}, z^{\ell}_{\text{noise}} \in \mathbb{R}^{c_\ell \times n_\ell}\) are concatenated along the channel axis
\begin{equation}
  u^{\ell} = \mathrm{Concat}(z_{\text{total}}^{\ell}, z_{\text{noise}}^{\ell}) \in \mathbb{R}^{2c_\ell \times n_\ell}~,
\end{equation}
forming a combined representation of the noisy and total content.

This tensor is processed by a correction block \(\phi^{\ell}\), defined as a two-layer perceptron implemented using sequential \(\mathrm{Conv1D}_1\) operations with \texttt{ReLU} activations
\begin{equation}
  \phi^{\ell}(u) = \mathrm{Conv1D}_1 \circ \mathrm{ReLU} \circ \mathrm{Conv1D}_1(u)~,
\end{equation}
where \(\mathrm{Conv1D}_1\) and \(\mathrm{Conv1D}_1\) are 1D convolutions of kernel size 1 with learnable weights \(W_1^{\ell} \in \mathbb{R}^{c' \times 2c_\ell \times 1}\), \(W_2^{\ell} \in \mathbb{R}^{c_\ell \times c' \times 1}\), biases \(b_1^{\ell} \in \mathbb{R}^{c'}\), \(b_2^{\ell} \in \mathbb{R}^{c_\ell}\).

The output is a learned correction map
\begin{equation}
  C_{\text{corr}}^{\ell}(u^{\ell}) = \phi^{\ell}(u^{\ell}) \in \mathbb{R}^{c_\ell \times n_\ell}~,
\end{equation}
which has the same shape as the original latent map.

The corrected signal latent is then computed as
\begin{equation}
  z_{\text{signal}}^{\ell} = z_{\text{total}}^{\ell} - z_{\text{noise}}^{\ell} + C_{\text{corr}}^{\ell}(u^{\ell})~,
\end{equation}
allowing the model to recover signal features that would be suppressed by direct subtraction alone.

The full set of corrected signal latents across all scales forms
\begin{equation}
  \tilde{z}_{\text{signal}} = \{ z_{\text{signal}}^{\ell} \}_{\ell=1}^{L}~,
\end{equation}
which is passed to the signal decoder for reconstruction of the estimated gravitational-wave background. This latent-level correction is crucial to retain faint or spatially correlated features that may otherwise be lost during aggressive denoising.


The implementation of the network architecture detailed above, including its core components like convolutional layers, activation functions, normalization, and tensor operations, was performed using the PyTorch deep learning framework~\cite{Paszke_2019_NeurIPS}.

\section{Training Algorithm Details}
\label{app:training_details}

In what follows, we describe 
the training algorithm used to optimize the model introduced in Section~\ref{subsec:training_methodology}. 
We outline the update rules, gradient handling techniques, and dynamic learning rate control mechanisms employed to ensure both convergence and numerical stability during training.

The MSMHAutoencoder's neurons weights \( \theta \) are updated via the \texttt{Adam} optimization algorithm~\cite{Adam}. At each step \( t \), the gradient of the total loss function \( \mathcal{L}_{\text{total}} \) is computed as \( g_t = \nabla_\theta \mathcal{L}_{\text{total}} \). Two moment estimates are 
 maintained and updated using exponential moving averages
\begin{equation}
\begin{aligned}
m_t &= \beta_1 m_{t-1} + (1 - \beta_1) g_t~, \\
v_t &= \beta_2 v_{t-1} + (1 - \beta_2) g_t^2~, \\
\hat{m}_t &= \frac{m_t}{1 - \beta_1^t}~, \\
\hat{v}_t &= \frac{v_t}{1 - \beta_2^t}~, \\
\theta_{t+1} &= \theta_t - \eta_t \frac{\hat{m}_t}{\sqrt{\hat{v}_t} + \epsilon_{\text{Adam}}}~,
\end{aligned}
\label{eq:app_adam_update}
\end{equation}
where \( \beta_1, \beta_2 \in [0, 1) \) are the decay rates for the first and second moments respectively, \( \epsilon_{\text{Adam}} \) is a small constant added for numerical stability, and \( \eta_t \) is the learning rate at step \( t \).

To avoid instability due to large gradients, we apply gradient clipping after backpropagation but before the optimizer update. If the L2 norm of the gradient vector exceeds a fixed threshold \( C_{\text{clip}} = 0.1 \), the gradient is rescaled
\begin{equation}
  g_t \leftarrow g_t \times \frac{C_{\text{clip}}}{\|g_t\|_2} \quad \text{if } \ \ \|g_t\|_2 > C_{\text{clip}}~.
\end{equation}
This ensures that the update step remains bounded, particularly when using mixed precision training.

The learning rate \( \eta_t \) is managed through a validation-aware adaptive scheduler based on the \texttt{ReduceLROnPlateau} ~\cite{Goodfellow-et-al-2016}. At the end of each epoch \( t \), a smoothed validation loss \( \bar{L}_{\text{val}}(t) \) is computed using a moving average over the last \( W = 5 \) epochs
\begin{equation}
\bar{L}_{\text{val}}(t) = \begin{cases}
\frac{1}{W} \sum\limits_{k=t-W+1}^{t} L_{\text{val}}(k) & \text{if } \ t \ge W \\
L_{\text{val}}(t) & \text{if } \ t < W~.
\end{cases}
\end{equation}
The best smoothed loss observed so far is stored as \( \bar{L}^*(t) = \min\limits_{1 \le k \le t} \bar{L}_{\text{val}}(k) \). If the validation loss has not improved over a patience interval of \( p = 10 \) epochs (i.e., \( \bar{L}_{\text{val}}(t) \ge \bar{L}^*(t - p) \)), the learning rate is reduced
\begin{equation}
\eta_{t+1} = \max(\gamma \times \eta_t~,\; \eta_{\min})~,
\end{equation}
where \( \gamma = 0.5 \) is the reduction factor and \( \eta_{\min} = 10^{-8} \) prevents the learning rate from vanishing entirely.

To prevent overfitting, early stopping is triggered when no meaningful improvement is detected over a long interval. Let \( \delta_{\text{stop}} = 10^{-5} \) denote the minimum required improvement in the validation loss. If this level of improvement is not achieved for \( P = 30 \) consecutive epochs, training is halted. Specifically, let \( T \) be the current epoch, and let \( \bar{L}^*(T) = \min_{1 \le k \le T} \bar{L}_{\text{val}}(k) \) be the best smoothed loss. If for all \( j \in [T - P + 1,\, T] \), we have
\begin{equation}
\bar{L}_{\text{val}}(j) \ge \bar{L}^*(T) - \delta_{\text{stop}}~,
\end{equation}
then training is terminated.

\newpage
\bibliography{ML_long}

\begin{thebibliography}{10}

\bibitem{LIGOScientific:2018mvr}
B.~P. Abbott et~al.
\newblock {GWTC-1: A Gravitational-Wave Transient Catalog of Compact Binary
  Mergers Observed by LIGO and Virgo during the First and Second Observing
  Runs}.
\newblock {\em Phys. Rev.}, X9(3):031040, 2019.

\bibitem{LIGOScientific:2020ibl}
R.~Abbott et~al.
\newblock {GWTC-2: Compact Binary Coalescences Observed by LIGO and Virgo
  During the First Half of the Third Observing Run}.
\newblock {\em Phys. Rev. X}, 11:021053, 2021.

\bibitem{GWTC-3-LIGOScientific:2021djp}
R.~Abbott et~al.
\newblock {GWTC-3: Compact Binary Coalescences Observed by LIGO and Virgo
  During the Second Part of the Third Observing Run}.
\newblock 11 2021.

\bibitem{Capote_2025}
E.~Capote et~al.
\newblock Advanced ligo detector performance in the fourth observing run.
\newblock {\em Physical Review D}, 111(6), March 2025.

\bibitem{Christensen:2018iqi}
Nelson Christensen.
\newblock {Stochastic Gravitational Wave Backgrounds}.
\newblock {\em Rept. Prog. Phys.}, 82(1):016903, 2019.

\bibitem{Caprini:2018mtu}
Chiara Caprini and Daniel~G. Figueroa.
\newblock {Cosmological Backgrounds of Gravitational Waves}.
\newblock {\em Class. Quant. Grav.}, 35(16):163001, 2018.

\bibitem{Kalogera:2021bya}
Vicky Kalogera et~al.
\newblock {The Next Generation Global Gravitational Wave Observatory: The
  Science Book}.
\newblock 11 2021.

\bibitem{LISACosmologyWorkingGroup:2022jok}
Pierre Auclair et~al.
\newblock {Cosmology with the Laser Interferometer Space Antenna}.
\newblock {\em Living Rev. Rel.}, 26(1):5, 2023.

\bibitem{Regimbau:2011rp}
Tania Regimbau.
\newblock {The astrophysical gravitational wave stochastic background}.
\newblock {\em Res. Astron. Astrophys.}, 11:369--390, 2011.

\bibitem{LIGOScientific:2021nrg}
R.~Abbott et~al.
\newblock {Constraints on Cosmic Strings Using Data from the Third Advanced
  LIGO\textendash{}Virgo Observing Run}.
\newblock {\em Phys. Rev. Lett.}, 126(24):241102, 2021.

\bibitem{Caldwell_2022}
Robert Caldwell, Yanou Cui, Huai-Ke Guo, Vuk Mandic, Alberto Mariotti,
  Jose~Miguel No, Michael~J. Ramsey-Musolf, Mairi Sakellariadou, Kuver Sinha,
  Lian-Tao Wang, Graham White, Yue Zhao, Haipeng An, Ligong Bian, Chiara
  Caprini, Sebastien Clesse, James~M. Cline, Giulia Cusin, Bartosz Fornal,
  Ryusuke Jinno, Benoit Laurent, Noam Levi, Kun-Feng Lyu, Mario Martinez,
  Andrew~L. Miller, Diego Redigolo, Claudia Scarlata, Alexander Sevrin, Barmak
  Shams~Es Haghi, Jing Shu, Xavier Siemens, Danièle~A. Steer, Raman Sundrum,
  Carlos Tamarit, David~J. Weir, Ke-Pan Xie, Feng-Wei Yang, and Siyi Zhou.
\newblock Detection of early-universe gravitational-wave signatures and
  fundamental physics.
\newblock {\em General Relativity and Gravitation}, 54(12), November 2022.

\bibitem{Blasi:2022ayo}
Simone Blasi, Alberto Mariotti, A\"aron Rase, Alexander Sevrin, and Kevin
  Turbang.
\newblock {Friction on ALP domain walls and gravitational waves}.
\newblock {\em JCAP}, 04:008, 2023.

\bibitem{Martinovic:2021hzy}
Katarina Martinovic, Charles Badger, Mairi Sakellariadou, and Vuk Mandic.
\newblock {Searching for parity violation with the LIGO-Virgo-KAGRA network}.
\newblock {\em Phys. Rev. D}, 104(8):L081101, 2021.

\bibitem{Badger:2022nwo}
Charles Badger et~al.
\newblock {Probing early Universe supercooled phase transitions with
  gravitational wave data}.
\newblock {\em Phys. Rev. D}, 107(2):023511, 2023.

\bibitem{Duval:2024jsg}
Hannah Duval, Sachiko Kuroyanagi, Alberto Mariotti, Alba Romero-Rodr\'\i{}guez,
  and Mairi Sakellariadou.
\newblock {Investigating cosmic histories with a stiff era through
  gravitational waves}.
\newblock {\em Phys. Rev. D}, 110(10):103503, 2024.

\bibitem{Badger:2024ekb}
Charles Badger, Hannah Duval, Tomohiro Fujita, Sachiko Kuroyanagi, Alba
  Romero-Rodr\'\i{}guez, and Mairi Sakellariadou.
\newblock {Detection prospects of gravitational waves from SU(2) axion
  inflation}.
\newblock {\em Phys. Rev. D}, 110(8):084063, 2024.

\bibitem{LIGOScientific:2017zlf}
Benjamin~P. Abbott et~al.
\newblock {GW170817: Implications for the Stochastic Gravitational-Wave
  Background from Compact Binary Coalescences}.
\newblock {\em Phys. Rev. Lett.}, 120(9):091101, 2018.

\bibitem{Christensen:1996da}
Nelson Christensen.
\newblock {Optimal detection strategies for measuring the stochastic
  gravitational radiation background with laser interferometric antennas}.
\newblock {\em Phys. Rev. D}, 55:448--454, 1997.

\bibitem{LIGOScientific:2016jlg}
Benjamin~P. Abbott et~al.
\newblock {Upper Limits on the Stochastic Gravitational-Wave Background from
  Advanced LIGO\textquoteright{}s First Observing Run}.
\newblock {\em Phys. Rev. Lett.}, 118(12):121101, 2017.
\newblock [Erratum: Phys.Rev.Lett. 119, 029901 (2017)].

\bibitem{LIGOScientific:2019vic}
B.~P. Abbott et~al.
\newblock {Search for the isotropic stochastic background using data from
  Advanced LIGO\textquoteright{}s second observing run}.
\newblock {\em Phys. Rev. D}, 100(6):061101, 2019.

\bibitem{KAGRA:2021kbb}
R.~Abbott et~al.
\newblock {Upper limits on the isotropic gravitational-wave background from
  Advanced LIGO and Advanced Virgo\textquoteright{}s third observing run}.
\newblock {\em Phys. Rev. D}, 104(2):022004, 2021.

\bibitem{Meyers:2020qrb}
Patrick~M. Meyers, Katarina Martinovic, Nelson Christensen, and Mairi
  Sakellariadou.
\newblock {Detecting a stochastic gravitational-wave background in the presence
  of correlated magnetic noise}.
\newblock {\em Phys. Rev. D}, 102(10):102005, 2020.

\bibitem{Parida_2016}
Abhishek Parida, Sanjit Mitra, and Sanjay Jhingan.
\newblock Component separation of a isotropic gravitational wave background.
\newblock {\em Journal of Cosmology and Astroparticle Physics},
  2016(04):024–024, April 2016.

\bibitem{Suresh_2021}
Jishnu Suresh, Deepali Agarwal, and Sanjit Mitra.
\newblock Jointly setting upper limits on multiple components of an anisotropic
  stochastic gravitational-wave background.
\newblock {\em Physical Review D}, 104(10), November 2021.

\bibitem{Boileau_2021}
Guillaume Boileau, Nelson Christensen, Renate Meyer, and Neil~J. Cornish.
\newblock Spectral separation of the stochastic gravitational-wave background
  for lisa: Observing both cosmological and astrophysical backgrounds.
\newblock {\em Physical Review D}, 103(10), May 2021.

\bibitem{Carleo:2019ptp}
Giuseppe Carleo, Ignacio Cirac, Kyle Cranmer, Laurent Daudet, Maria Schuld,
  Naftali Tishby, Leslie Vogt-Maranto, and Lenka Zdeborov\'a.
\newblock {Machine learning and the physical sciences}.
\newblock {\em Rev. Mod. Phys.}, 91(4):045002, 2019.

\bibitem{ntampaka2019deep}
Michelle Ntampaka, J~ZuHone, D~Eisenstein, D~Nagai, A~Vikhlinin, L~Hernquist,
  Federico Marinacci, Dylan Nelson, R{\"u}diger Pakmor, Annalisa Pillepich,
  et~al.
\newblock A deep learning approach to galaxy cluster x-ray masses.
\newblock {\em The Astrophysical Journal}, 876(1):82, 2019.

\bibitem{George:2017pmj}
Daniel George and E.~A. Huerta.
\newblock {Deep Learning for Real-time Gravitational Wave Detection and
  Parameter Estimation: Results with Advanced LIGO Data}.
\newblock {\em Phys. Lett. B}, 778:64--70, 2018.

\bibitem{Schafer:2022dxv}
Marlin~B. Sch\"afer et~al.
\newblock {First machine learning gravitational-wave search mock data
  challenge}.
\newblock {\em Phys. Rev. D}, 107(2):023021, 2023.

\bibitem{Chua:2019wwt}
Alvin J.~K. Chua and Michele Vallisneri.
\newblock {Learning Bayesian posteriors with neural networks for
  gravitational-wave inference}.
\newblock {\em Phys. Rev. Lett.}, 124(4):041102, 2020.

\bibitem{Gabbard:2019rde}
Hunter Gabbard, Chris Messenger, Ik~Siong Heng, Francesco Tonolini, and
  Roderick Murray-Smith.
\newblock {Bayesian parameter estimation using conditional variational
  autoencoders for gravitational-wave astronomy}.
\newblock {\em Nature Phys.}, 18(1):112--117, 2022.

\bibitem{Cuoco:2024cdk}
Elena Cuoco, Marco Cavagli\`a, Ik~Siong Heng, David Keitel, and Christopher
  Messenger.
\newblock {Applications of machine learning in gravitational-wave research with
  current interferometric detectors}.
\newblock {\em Living Rev. Rel.}, 28(1):2, 2025.

\bibitem{2005PhRvD..72h4001B}
Alessandra {Buonanno}, G{\"u}nter {Sigl}, Georg~G. {Raffelt}, Hans-Thomas
  {Janka}, and Ewald {M{\"u}ller}.
\newblock {Stochastic gravitational-wave background from cosmological
  supernovae}.
\newblock {\em \prd}, 72(8):084001, October 2005.

\bibitem{Auclair:2019wcv}
Pierre Auclair et~al.
\newblock {Probing the gravitational wave background from cosmic strings with
  LISA}.
\newblock {\em JCAP}, 04:034, 2020.

\bibitem{Thrane:2014yza}
E.~Thrane, N.~Christensen, R.~M.~S. Schofield, and A.~Effler.
\newblock {Correlated noise in networks of gravitational-wave detectors:
  subtraction and mitigation}.
\newblock {\em Phys. Rev. D}, 90(2):023013, 2014.

\bibitem{Martinovic:2020hru}
Katarina Martinovic, Patrick~M. Meyers, Mairi Sakellariadou, and Nelson
  Christensen.
\newblock {Simultaneous estimation of astrophysical and cosmological stochastic
  gravitational-wave backgrounds with terrestrial detectors}.
\newblock {\em Phys. Rev. D}, 103(4):043023, 2021.

\bibitem{2017PhRvL.118o1105R}
T.~{Regimbau}, M.~{Evans}, N.~{Christensen}, E.~{Katsavounidis},
  B.~{Sathyaprakash}, and S.~{Vitale}.
\newblock {Digging Deeper: Observing Primordial Gravitational Waves below the
  Binary-Black-Hole-Produced Stochastic Background}.
\newblock {\em \prl}, 118(15):151105, April 2017.

\bibitem{2020PhRvD.102b4051S}
Surabhi {Sachdev}, Tania {Regimbau}, and B.~S. {Sathyaprakash}.
\newblock {Subtracting compact binary foreground sources to reveal primordial
  gravitational-wave backgrounds}.
\newblock {\em \prd}, 102(2):024051, July 2020.

\bibitem{2020PhRvD.102f3009S}
Ashish {Sharma} and Jan {Harms}.
\newblock {Searching for cosmological gravitational-wave backgrounds with
  third-generation detectors in the presence of an astrophysical foreground}.
\newblock {\em \prd}, 102(6):063009, September 2020.

\bibitem{2023PhRvD.108f4040Z}
Bei {Zhou}, Luca {Reali}, Emanuele {Berti}, Mesut {{\c{c}}al{\i}{\c{s}}kan},
  Cyril {Creque-Sarbinowski}, Marc {Kamionkowski}, and B.~S. {Sathyaprakash}.
\newblock {Subtracting compact binary foregrounds to search for subdominant
  gravitational-wave backgrounds in next-generation ground-based
  observatories}.
\newblock {\em \prd}, 108(6):064040, September 2023.

\bibitem{Zhong:2024dss}
Haowen Zhong, Bei Zhou, Luca Reali, Emanuele Berti, and Vuk Mandic.
\newblock {Searching for cosmological stochastic backgrounds by notching out
  resolvable compact binary foregrounds with next-generation gravitational-wave
  detectors}.
\newblock {\em Phys. Rev. D}, 110(6):064047, 2024.

\bibitem{Biscoveanu:2020gds}
Sylvia Biscoveanu, Colm Talbot, Eric Thrane, and Rory Smith.
\newblock {Measuring the primordial gravitational-wave background in the
  presence of astrophysical foregrounds}.
\newblock {\em Phys. Rev. Lett.}, 125:241101, 2020.

\bibitem{Bourlard1988}
H.~Bourlard and Y.~Kamp.
\newblock Auto-association by multilayer perceptrons and singular value
  decomposition.
\newblock {\em Biological Cybernetics}, 59(4-5):291--294, 1988.

\bibitem{Hinton2006}
G.~E. Hinton and R.~R. Salakhutdinov.
\newblock Reducing the dimensionality of data with neural networks.
\newblock {\em Science}, 313(5786):504--507, 2006.

\bibitem{Goodfellow-et-al-2016}
Ian Goodfellow, Yoshua Bengio, and Aaron Courville.
\newblock {\em Deep Learning}.
\newblock MIT Press, 2016.

\bibitem{Rumelhart1986}
David~E. Rumelhart, Geoffrey~E. Hinton, and Ronald~J. Williams.
\newblock Learning representations by back-propagating errors.
\newblock {\em Nature}, 323(6088):533--536, 1986.

\bibitem{Adam}
Diederik~P. Kingma and Jimmy Ba.
\newblock Adam: A method for stochastic optimization, 2017.

\bibitem{Prechelt_1998_EarlyStopping}
Lutz Prechelt.
\newblock Early stopping -- but when?
\newblock In Gr\'{e}goire Montavon, Genevi\`{e}ve~B. Orr, and Klaus-Robert
  M\"{u}ller, editors, {\em Neural Networks: Tricks of the Trade}, pages
  55--69. Springer Berlin Heidelberg, 1998.

\bibitem{T1800042:2018}
Lisa Barsotti, Lee McCuller, Peter Fritschel, and Matthew Evans.
\newblock {The A+ design curve}.
\newblock {\em https://dcc.ligo.org/T1800042-v5/public}, 2018.

\bibitem{lalsuite}
{LIGO Scientific Collaboration}, {Virgo Collaboration}, and {KAGRA
  Collaboration}.
\newblock {LVK} {A}lgorithm {L}ibrary - {LALS}uite.
\newblock Free software (GPL), 2018.

\bibitem{Mapelli:2021syv}
Michela Mapelli et~al.
\newblock {Hierarchical black hole mergers in young, globular and nuclear star
  clusters: the effect of metallicity, spin and cluster properties}.
\newblock {\em Mon. Not. Roy. Astron. Soc.}, 505(1):339--358, 2021.

\bibitem{Santoliquido:2020axb}
Filippo Santoliquido, Michela Mapelli, Nicola Giacobbo, Yann Bouffanais, and
  M.~Celeste Artale.
\newblock {The cosmic merger rate density of compact objects: impact of star
  formation, metallicity, initial mass function and binary evolution}.
\newblock {\em Mon. Not. Roy. Astron. Soc.}, 502(4):4877--4889, 2021.

\bibitem{Dietrich:2019kaq}
Tim Dietrich, Anuradha Samajdar, Sebastian Khan, Nathan~K. Johnson-McDaniel,
  Reetika Dudi, and Wolfgang Tichy.
\newblock {Improving the NRTidal model for binary neutron star systems}.
\newblock {\em Phys. Rev. D}, 100(4):044003, 2019.

\bibitem{Pratten:2020ceb}
Geraint Pratten et~al.
\newblock {Computationally efficient models for the dominant and subdominant
  harmonic modes of precessing binary black holes}.
\newblock {\em Phys. Rev. D}, 103(10):104056, 2021.

\bibitem{regimbau:2025xxx}
Tania Regimbau and Jishnu Suresh.
\newblock A mock data challenge for next generation detectors.
\newblock 2025.

\bibitem{2003CQGra..20S.677B}
Sukanta {Bose}, Bruce {Allen}, Michael {Landry}, Albert {Lazzarini}, Isabel
  {Leonor}, Szabolcs {Marka}, Tania {Regimbau}, Joseph {Romano}, Peter
  {Shawhan}, Daniel {Sigg}, and John {Whelan}.
\newblock {Towards the first search for a stochastic background in LIGO data:
  applications of signal simulations}.
\newblock {\em Classical and Quantum Gravity}, 20(17):S677--S687, September
  2003.

\bibitem{2007CQGra..24S.639C}
Giancarlo {Cella}, Carlo~Nicola {Colacino}, Elena {Cuoco}, Angela {Di
  Virgilio}, Tania {Regimbau}, Emma~L. {Robinson}, and John~T. {Whelan}.
\newblock {Prospects for stochastic background searches using Virgo and LSC
  interferometers}.
\newblock {\em Classical and Quantum Gravity}, 24(19):S639--S648, October 2007.

\bibitem{emcee}
D.~{Foreman-Mackey}, D.~W. {Hogg}, D.~{Lang}, and J.~{Goodman}.
\newblock {emcee: The MCMC Hammer}.
\newblock {\em Publications of the Astronomical Society of the Pacific},
  125(925):306--312, 2013.

\bibitem{dynesty}
Joshua~S. Speagle.
\newblock {D}ynesty: a dynamic nested sampling package for estimating bayesian
  posteriors and evidences.
\newblock {\em Monthly Notices of the Royal Astronomical Society},
  493(3):3132--3158, January 2020.

\bibitem{jeffreys1998theory}
Harold Jeffreys.
\newblock {\em The theory of probability}.
\newblock OuP Oxford, 1998.

\bibitem{1999PhRvD..59j2001A}
Bruce {Allen} and Joseph~D. {Romano}.
\newblock {Detecting a stochastic background of gravitational radiation: Signal
  processing strategies and sensitivities}.
\newblock {\em \prd}, 59(10):102001, May 1999.

\bibitem{2013PhRvD..88l4032T}
Eric {Thrane} and Joseph~D. {Romano}.
\newblock {Sensitivity curves for searches for gravitational-wave backgrounds}.
\newblock {\em \prd}, 88(12):124032, December 2013.

\bibitem{2022Galax..10...34R}
Arianna~I. {Renzini}, Boris {Goncharov}, Alexander~C. {Jenkins}, and Patrick~M.
  {Meyers}.
\newblock {Stochastic Gravitational-Wave Backgrounds: Current Detection Efforts
  and Future Prospects}.
\newblock {\em Galaxies}, 10(1):34, February 2022.

\bibitem{Thrane2013}
Thrane E. and Romano J.
\newblock Sensitivity curves for searches for gravitational-wave backgrounds.
\newblock Technical Report LIGO-P1300115, LIGO Document Control Center, 2013.
\newblock \url{https://dcc.ligo.org/LIGO-P1300115-v6/public}.

\bibitem{Szegedy_2015_CVPR}
Christian Szegedy, Wei Liu, Yangqing Jia, Pierre Sermanet, Scott Reed, Dragomir
  Anguelov, Dumitru Erhan, Vincent Vanhoucke, and Andrew Rabinovich.
\newblock Going deeper with convolutions.
\newblock In {\em Proceedings of the IEEE Conference on Computer Vision and
  Pattern Recognition (CVPR)}, pages 1--9, June 2015.

\bibitem{Paszke_2019_NeurIPS}
Adam Paszke, Sam Gross, Francisco Massa, Adam Lerer, James Bradbury, Gregory
  Chanan, Trevor Killeen, Zeming Lin, Natalia Gimelshein, Luca Antiga, Alban
  Desmaison, Andreas Kopf, Edward Yang, Zachary DeVito, Martin Raison, Alykhan
  Tejani, Sasank Chilamkurthy, Benoit Steiner, Lu~Fang, Junjie Bai, and Soumith
  Chintala.
\newblock Pytorch: An imperative style, high-performance deep learning library.
\newblock In H.~Wallach, H.~Larochelle, A.~Beygelzimer, F.~d'Alch\'{e} Buc,
  E.~Fox, and R.~Garnett, editors, {\em Advances in Neural Information
  Processing Systems 32 (NeurIPS 2019)}, pages 8026--8037. Curran Associates,
  Inc., 2019.

\bibitem{Ronneberger_2015_MICCAI}
Olaf Ronneberger, Philipp Fischer, and Thomas Brox.
\newblock {U-Net:} convolutional networks for biomedical image segmentation.
\newblock In {\em Medical Image Computing and Computer-Assisted Intervention --
  MICCAI 2015}, volume 9351 of {\em Lecture Notes in Computer Science}, pages
  234--241. Springer International Publishing, 2015.

\bibitem{Zeiler_2014_ECCV}
Matthew~D. Zeiler and Rob Fergus.
\newblock Visualizing and understanding convolutional networks.
\newblock In {\em Computer Vision -- ECCV 2014}, volume 8689 of {\em Lecture
  Notes in Computer Science}, pages 818--833. Springer International
  Publishing, 2014.

\bibitem{Parker_1983_TMI}
J.~Anthony Parker, Robert~V. Kenyon, and Donald~E. Troxel.
\newblock Comparison of interpolating methods for image resampling.
\newblock {\em IEEE Transactions on Medical Imaging}, 2(1):31--39, 1983.

\bibitem{Glorot_2011_AISTATS}
Xavier Glorot, Antoine Bordes, and Yoshua Bengio.
\newblock Deep sparse rectifier neural networks.
\newblock In {\em Proceedings of the Fourteenth International Conference on
  Artificial Intelligence and Statistics (AISTATS)}, volume~15 of {\em
  Proceedings of Machine Learning Research}, pages 315--323. PMLR, 2011.

\end{thebibliography}
\end{document}